\documentclass[a4paper,12pt]{article}%amsart}
\usepackage[utf8x]{inputenc}

\usepackage{a4wide}
\usepackage{amssymb}
\usepackage{amsmath}
\usepackage{amsthm}
\usepackage[mathscr]{euscript}
\usepackage[colorlinks=true]{hyperref}
\usepackage[usenames,dvipsnames]{xcolor}
\usepackage{todonotes}
\usepackage{appendix}

\numberwithin{equation}{section}

\theoremstyle{plain}

\newtheorem{definition}{Definition}[section]

\newtheorem{theorem}[definition]{Theorem}
\newtheorem{proposition}[definition]{Proposition}
\newtheorem{lemma}[definition]{Lemma}

\theoremstyle{remark}

\usepackage[shortlabels]{enumitem}
\setlist[itemize]{itemsep=1pt,topsep=2pt}
\usepackage[nottoc,notlot,notlof]{tocbibind}
\usepackage{footmisc}

\newcommand{\Ric}{\mathrm{Ric}}
\newcommand{\tr}{\mathrm{tr}}
\newcommand{\id}{\mathrm{id}}
\newcommand{\pa}{\partial}
\newcommand{\Reals}{{\mathbb R}}
\newcommand{\N}{{\mathbb N}}

\newcommand{\eps}{\varepsilon}

\usepackage[xcolor]{changebar}
\cbcolor{blue}
\newcommand{\X}{\mathfrak{X}}

\newcommand{\sse}{\subseteq}

\newcommand{\D}{\mathcal{D}}
\newcommand{\diag}{\mathrm{diag}}

\newcommand{\Dpk}{\mathcal{D}'{}^{(k)}}
\newcommand{\Dpo}{\mathcal{D}'{}^{(1)}}

\newcommand{\ltl}{L^2_{\mathrm{loc}}}
\newcommand{\lpl}{L^p_{\mathrm{loc}}}

\newcommand{\Om}{\Omega}
\begin{document}

\title{The singularity theorems of General Relativity\\[.2cm] and their low regularity extensions}
\author{Roland Steinbauer\\
\small{University of Vienna, Faculty of Mathematics}\\ 
\small{Oskar-Morgenstern-Platz 1, A-1090 Wien, Austria}\\ 
\small{roland.steinbauer@univie.ac.at}}

\maketitle
\thispagestyle{empty}

\abstract{
On the occasion of Sir Roger Penrose's 2020 Nobel Prize in Physics, we review the singularity theorems of General Relativity, as well as their recent extension to Lorentzian metrics of low regularity. The latter is motivated by the quest to explore the nature of the singularities predicted by the classical theorems. Aiming at the more mathematically minded reader, we give a pedagogical introduction to the classical theorems with an emphasis on the analytical side of the arguments. We especially concentrate on focusing results for causal geodesics under appropriate geometric and initial conditions, in the smooth and in the low regularity case. The latter comprise the main technical advance that leads to the proofs of $C^1$-singularity theorems via a regularisation approach that allows to deal with the distributional curvature. We close with an overview on related lines of research and a future outlook.
}
\bigskip

\bigskip

%83C75: Space-time singularities, cosmic censorship, etc.
%53B30: Lorentz metrics, indefinite metrics
%83-02: Research exposition (monographs, survey articles)

\pagebreak
 \tableofcontents

\section{Introduction}\label{sec1}

The singularity theorems of General Relativity (GR) are commonly counted among the 20th century milestones in mathematical physics. They comprise a body of rigorous results in Lorentzian differential geometry that, under physically reasonable conditions, imply the occurrence of a ``singularity'' in the sense of causal geodesic incompleteness of the spacetime manifold. At the time of its appearance, the first singularity theorem proved by Roger Penrose in \cite{Pen:65} came as a surprise to the community, debunking the widely held belief that the singularities encountered in many of the exact solutions of GR were a mere artefact of their high degree of symmetry. It has even been argued e.g.\ in \cite[Sec.\ 15.1]{Sen:12} that Penrose's theorem, that appeared exactly 50 years after the birth of GR, was the first true post-Einsteinian contribution in the sense that it was not foreseen by its founding father. And indeed, it is this paper that won Roger Penrose (one half of) the 2020 Nobel Prize in Physics. Despite its brevity of less than three pages, it not only shaped the way we now primarily think of singularities---via \emph{geodesic incompleteness}---but also put forward the fundamental new idea of a \emph{trapped surface,} which has stimulated mathematical GR up to the present day. The theorem itself roughly says that, in a situation that physically amounts to the formation of a black hole---intuitively a region in spacetime of so strong gravity that not even light can escape---there necessarily exists an incomplete null geodesic, i.e., a light ray that ends suddenly\footnote{It has to be pointed out, however, that the theorem---quite contrary to its widespread folklore transcript---does \emph{not} say that black holes form in gravitational collapse, see e.g.\cite[Sec.\ 4a]{Sen:21} for details. In fact, it is here where Penrose's cosmic censorship hypotheses originate, for more details see e.g.\ \cite{Lan:21}.}. 

A short time later, Stephen Hawking realised that one could apply a similar reasoning to an everywhere expanding universe and showed that in such a spacetime there has to be a timelike geodesic that is past incomplete. This theorem, known as the Hawking singularity theorem \cite{Haw:67}, is generally thought of as mathematical evidence for the occurrence of a \emph{big bang}. Then, during the next couple of years, Robert Geroch, George Ellis and others helped to shape a body of results which we now call the classical singularity theorems---for some historical details see e.g.\ \cite[Sec.\ 5]{SG:15}. Notably, in 1970 Hawking and Penrose, in their only joint paper \cite{HP:70}, proved the most refined of these results which is known as the Hawking-Penrose singularity theorem. 
\medskip

Over the decades the singularity theorems have not only become an integral part of GR, but are still an area of active research, cf.\ e.g.\ \cite[Sec. 8]{SG:15}. There is, of course, an extensive textbook coverage available and we only  mention the early classics \cite{Pen:72b,HE:73}, and the more mathematically oriented standard accounts \cite{BEE:96,ONe:83,Cla:93,Kri:99}, as well as the review articles \cite{Sen:98,SG:15,MS:15,Sen:21,Daf:21,Lan:21}, the latest three of which have appeared on the occasion of the 2020 Nobel Prize.

In this work we offer a guided tour for the mathematically oriented reader to the classical singularity theorems and to their recent extensions for Lorentzian metrics of low regularity. The study of the latter is motivated by the discussion of the ``character of the singularities'' (cf.\ \cite [Sec.\ 8.4]{HE:73}) predicted by the theorems,  which we will recall after presenting the classical results.

In some more detail we briefly collect the necessary preliminaries on GR and Lorentzian geometry in Section \ref{sec:gr}, in a way especially suited to the thread of this account. In Section \ref{sec:thms} we discuss the main classical theorems, namely the one of Penrose, the one of Hawking, and the one of Hawking-Penrose, together with the main arguments that enter their proofs. In particular, we will discuss the focusing of geodesics under certain curvature and other geometric conditions (Subsection \ref{sec:3.2}) as well as the relevant parts of Lorentzian causality theory (Subsection \ref{sec:3.4}). Then, in a brief analysis of the conditions and statements of the theorems in Subsection \ref{sec:4.1}, we motivate the quest for their low regularity extensions which are the topic of Section \ref{sec:lr}. Here we present an overview of these results which were obtained in the last couple of years, and, again provide some key insights on the techniques and arguments employed in their proofs. The main thread is laid out in Subsections \ref{sec:4.2}--\ref{sec:4.6} and \ref{sec:4.10}, while the remaining rather technical subsections are directed towards the more initiated reader. Finally, in Section \ref{sec:5} we provide a summary, some conclusions, and an outlook to current and future lines of research in the direction at hand.

\section{Preliminaries: General Relativity, geometry, singularities}\label{sec:gr}

Here we first collect in a nutshell the basics of GR, Albert Einsteins theory of space, time, gravitation, and matter. Then we turn to its geometric foundations, that is Lorentzian differential geometry and review the necessary preliminaries for our approach to the singularity theorems. Finally, we briefly discuss the very notion of a singular spacetime in GR.

\subsection{Some basics of General Relativity}
The stage of the theory is an $n$-dimensional\footnote{In classical GR, of course, $n=4$. But since the bulk of the theory---as far as this presentation is concerned---does not depend on this restriction, we keep the dimension general and set $n\geq 3$.} \emph{Lorentzian manifold} $(M,g)$ where the metric $g$ is a symmetric, non-degenerate $(0,2)$-tensor field, assigning to any point $p$ in the smooth manifold\footnote{We assume all manifolds to be Hausdorff, second countable, and connected.}  $M$ a scalar product $g_p$ on the tangent space $T_pM$ at $p$. This scalar product, however, is not positive definite, but has signature $(-1,+1,\dots,+1)$, which gives rise to the following distinction of tangent vectors: We call $v\in T_pM$ \emph{spacelike (timelike)} if $\langle v,v\rangle \equiv g_p(v,v)>0$ ($<0$). We call it \emph{null} if $\langle v,v\rangle=0$ but $v\not=0$\footnote{By convention the zero vector is spacelike.} and \emph{causal} if it is timelike or null. These notions extend naturally to vector fields and, via their tangent vectors, to sufficiently smooth curves. 

Given the existence of a timelike vector field on $M$ we can use it to define a \emph{time orientation}, i.e.\ a smooth choice of one of the two connected components of the set of causal vectors at any $p\in M$, called the future cone. \emph{A spacetime} then is a smooth Lorentzian manifold $(M,g)$ together with a time orientation. 

In this note we will also be interested in non-smooth spacetimes, that is smooth Lorentzian manifolds with a time oriented  metric of regularity below $C^2$, i.e., twice continuously differentiable. In fact, it makes sense to regard such metrics as being of \emph{low regularity} since the bulk of classical smooth Lorentzian geometry extends verbatim to $C^2$-metrics. We will in particular be interested in metrics of regularity $C^{1,1}$, i.e., metrics that are continuously differentiable ($C^1$) and, in addition, have first derivatives that are locally Lipschitz continuous\footnote{In the physical literature this class is often denoted by $C^{2-}$.}, as well as in merely $C^1$-  or $C^0$-metrics. We will always assume the time orientation to be induced by a \emph{smooth} timelike vector field and we do not reduce the regularity of the differential structure itself, since this is no loss of generality: Any $C^k$-manifold ($k\geq 1$) possesses a unique smooth structure compatible with the given $C^k$-structure. 
\medskip

At the heart of GR are the field equations which connect the curvature of spacetime to the matter and energy it contains. To make this statement precise, recall the \emph{Riemannian curvature tensor} of $(M,g)$ 
\begin{equation}\label{eq:riem}
  R(X,Y)Z=[\nabla_X,\nabla_Y]Z-\nabla_{[X,Y]}{Z},
\end{equation}
where $X, Y, Z\in {\mathfrak X}(M)=\Gamma(M,TM)$ are smooth vector fields and $\nabla$ denotes the Levi-Civita connection of $g$. Then we have the \emph{Ricci tensor} 
\begin{equation}\label{eq:ric}
 \Ric(X,Y)=\sum_{i=1}^n\langle E_i,E_i\rangle\langle R(E_i,X)Y,E_i\rangle,
\end{equation}
where here and in the following $(E_i)_{i=1}^n$ denotes a (local) orthonormal frame field.
%and $(e_i)_{i=1}^n$  denotes orthonormal frames in individual tangent spaces $T_pM$. 
Finally, writing $S$ for the \emph{curvature scalar}, i.e., the contraction of the Ricci tensor, we arrive at the \emph{Einstein equations}
\begin{equation}\label{eq:ees}
 \mathrm{ Ric}-\frac{1}{2}S\,g+\Lambda\,g\ =\ 8\,\pi\, T.
\end{equation}
Here $T$ is the \emph{energy momentum tensor} which encodes information on the matter and energy content of spacetime. We avoid a discussion of $T$ since we are mostly interested in the vacuum case, i.e., when there is no matter or energy and we have $T=0$. Also we will generally assume the \emph{cosmological constant} to vanish and set $\Lambda=0$.

The simplest vacuum solution of \eqref{eq:ees}, of course, is flat \emph{Minkowski space}, i.e., $\Reals^n$ with $g=\mathrm{diag}(-1,1,\dots,1)$, which is the stage of the special theory of relativity. Much more interesting is already the famous \emph{Schwarzschild metric} which gives the entire one-parameter family of spherically symmetric vacuum solutions of \eqref{eq:ees}. It models spacetime outside a spherically symmetric body of mass $m$ but the \emph{Kruskal extension} also is the basic model of a static black hole.

\medskip

Given these foundations, GR over its first decades was developed mainly as a geometric theory and a large zoo of solutions to the field equations were explicitly found and studied. This branch of GR known as \emph{exact solutions} (see e.g.\ \cite{SK:03,GP:09}) was complemented by more \emph{analytic approaches} that came into gear with the celebrated local existence result for the Einstein equations by Yvonne Choquet-Bruhat \cite{CB:52}.
During the past decades the PDE aspect of the theory has become more and more prevailing,  producing many remarkable existence and stability results, see e.g.\ \cite{Rin:15} for an overview. The past decades have also seen a growing impact of \emph{numerical methods} in GR (see e.g.\ \cite{BS:10}) and it finally was a combination of analytical and numerical methods that provided the theoretical background for the celebrated first direct observation of gravitational waves in 2015 \cite{GW:15}, see e.g.\ \cite{BGN:17}.
\medskip

Returning to the foundations of GR we proceed with a famous quote by John Wheeler cf.\ \cite[p.\ 5]{MTW:73}, in which he concisely describes its core:
\begin{quote}
Space tells matter how to move. Matter tells space how to curve. 
\end{quote}
While the latter statement refers to the field equations \eqref{eq:ees} introduced above, we now turn to the first one. Indeed matter, more precisely test particles of negligible (rest) mass and inner structure, often called (freely falling) observers, move along \emph{timelike geodesics}, while light rays trace out \emph{null geodesics} in spacetime. Collecting the basic facts on geodesics in Lorentzian manifolds we now begin our brief primer on the geometrical background of GR.

\subsection{Some basics of Lorentzian geometry}

Geodesics are curves $\gamma:I\to M$ defined on an interval $I$ that are self parallel, i.e., their tangents $\dot\gamma$ satisfy $\nabla_{\dot\gamma}\dot\gamma=0$. The \emph{geodesic equation} written in some local chart $(U,x^i=(x^0,\dots,x^{n-1}))$ takes the form 
\begin{equation}\label{eq:geo}
 \ddot x^i+\Gamma^i_{\,jk}\,\dot x^j\,\dot x^k = 0,
\end{equation}
with the Einstein summation convention in effect.
Here the Christoffel symbols are $\Gamma^i_{\,jk}=\frac{1}{2}g^{il}(\pa_k g_{lj}+\pa_jg_{kl}-\pa_lg_{jk})$, where $g_{jk}$ and $g^{jk}$ denote the coefficients of the spacetime metric and its inverse, respectively. Also we have followed the common habit to denote the components $x^i\circ\gamma$ of $\gamma$ simply by $x^i$.
Equation \eqref{eq:geo} is a second-order, non-linear ODE, hence given any point $p\in M$ and any tangent vector $v\in T_pM$ there is a unique geodesic $\gamma_v$ defined on a maximal half-open interval $I_v:=[0,\beta)$ assuming this data, i.e., $\gamma_v(0)=p$, and $\dot\gamma_v(0)=v$. We say that $\gamma_v$ is \emph{complete} if $\beta=\infty$. More generally, we call any geodesic $\gamma:I\to M$ \emph{extendible} if there exists a geodesic $\tilde\gamma:J\to M$ with $J\supsetneq I$ and $\tilde{\gamma}\mid_{I}=\gamma $. Otherwise $\gamma$ is called \emph{inextendible}. Finally, $\gamma$ is called \emph{complete} if it may be extended to all values of its (or any other affine) parameter, that is $\gamma:{\mathbb R} \to M$, and we call $M$ itself \emph{geodesically complete} it this is true for all geodesics.

Next we discuss the \emph{exponential map} at $p$ defined via 
\begin{equation}
 \exp_p:\  T_pM\supseteq{\cal D}\ni v\mapsto \exp_p(v):=\gamma_v(1) \in M,
\end{equation}
where the domain is ${\cal D}=\{v\in T_pM:\, 1\in I_v\}$. It is a basic fact that $\exp_p$ for any $p$ is a diffeomorphism from some open zero neighbourhood ${\cal U}\subseteq T_pM$ onto $U=\exp_p({\cal U})$ and we call $U$ a \emph{normal neighbourhood} of $p$ if ${\cal U}$ is star shaped. Any point $q$ in a normal neighbourhood $U$ of $p$ is connected to $p$ by a unique \emph{radial geodesic} $\gamma:[0,1]\to  U$ for which we have $\dot\gamma(0)=\exp_p^{-1}(q)$. A neighbourhood $U$ is called \emph{(geodesically) convex} if it is normal for all its points,  which implies the existence of a unique geodesic in $U$ between any pair of its points. Finally, every $p\in M$ possesses a base of convex neighbourhoods.
\medskip

Up to this point there do not occur any differences between Riemannian and Lorentzian manifolds. However, the Gauss Lemma, which states that the exponential map is a radial isometry, has quite different consequences in the two cases. While in the Riemannian case it implies that radial geodesics minimise the Riemannian distance in convex neighbourhoods, it here leads to radial causal geodesics \emph{maximising} the \emph{Lorentzian distance} (sometimes called the time separation function). 

To explain this statement in some detail, we need to introduce some basic notions from \emph{causality theory}, i.e., the theory of futures and pasts of points in a spacetime. For more details see \cite{MS:08} and the authoritative source \cite{Min:19a}. For $p,q\in M$ we write $p\ll q$ if there is a future directed timelike curve\footnote{While most textbooks base causality theory on piecewise smooth curves it has turned out to be more economically \cite{Chr:11,Min:19a} to use locally Lipschitz curves. Hence for us a curve is timelike (causal, null, future or past directed) if it is locally Lipschitz and $\dot{\gamma}(t)$,
which exists almost everywhere by Rademacher's theorem, is timelike (causal,
null, future or past directed) almost everywhere.}
from $p$ to $q$, and $q\leq q$ if there is a future directed causal curve from $p$ to $q$ or if $p=q$. We then define the \emph{chronological} and \emph{causal future} of $p$ via
\begin{equation}
 I^+(p)=\{q\in M:\ p\ll q\}, \quad\mbox{and}\quad J^+(p)=\{q\in M:\ p\leq q\},
\end{equation}
respectively. Analogously one defines the \emph{chronological} and \emph{causal past} $I^-(p)$ and $J^-(p)$ of a point, and for a set $A\subseteq M$ one defines $I^\pm(A)=\cup_{p\in A}I^\pm(p)$ and analogously for $J^\pm(A)$. It is a fundamental fact that $I^\pm(p)$ is open but $J^\pm(p)$ need neither be closed nor open. Also the so-called \emph{push-up principle} holds, which can bee seen as an improved transitivity of the causality relations: if $p\ll q$ and $q\leq r$ (or if $p\leq q$ and $q\ll r$), then $p\ll r$, and the name refers to the fact that a curve that has a timelike and a null part can be ``pushed up'' to give an overall timelike curve.

If $\gamma:I\to M$ is a (sufficiently smooth) curve, its  length is defined as $L(\gamma)=\int_I\sqrt{\mid\langle \dot\gamma(t),\dot\gamma(t)\rangle\mid}\,dt$. Given two timelike related points $p\ll q$ there are always arbitrarily short future directed timelike curves connecting them---just take curves arbitrarily close to a piecewise null zig-zag curve, which always has vanishing length. On the other hand, if $p$ and $q$ lie in a convex neighbourhood $U$ then the unique future directed timelike radial geodesic between them is the \emph{longest} curve in $U$ from $p$ to $q$. So it makes sense to define the \emph{Lorentzian distance} on $M\times M$ by
\begin{equation}
 d(p,q)=\begin{cases} \sup L(\gamma) &\mbox{if}\ q\in J^+(p)\\0 &\mbox{else,}\end{cases}
\end{equation}
where the $\sup$ runs over all future directed causal curves $\gamma$ from $p$ to $q$. Contrary to its Riemannian sister, the Lorentzian distance is not symmetric and it satisfies the \emph{reverse triangle inequality},
\begin{equation}
 d(p,q)\geq d(p,r)+d(r,q)\quad\mbox{for all $p\leq r\leq q$,}
\end{equation}
i.e., detours make curves shorter rather than longer. Also the $\sup$ is not finite if the causality of the spacetime behaves badly, e.g.\ if there are closed timelike curves.
\medskip

There are even more striking new phenomena in Lorentzian geometry when contrasted with the Riemannian case. Recall that the Riemannian distance $d:M\times M\to\Reals$ given by $d(p,q)=\inf\{L(\gamma)\}$ (where the $\inf$ runs over all curves $\gamma$ connecting $p$ and $q$) is a continuous metric that induces the manifold topology. Moreover, the \emph{Hopf-Rinow theorem} asserts that a Riemannian manifold $M$ is complete as a metric space iff it is geodesically complete 
%(i.e., all geodesics are defined for all values of an affine parameter) 
and iff $M$ is geodesically connected, i.e., every pair of points can be joined by a minimising geodesic $\gamma$, that is $L(\gamma)=d(p,q)$. 

In the Lorentzian case, as discussed above, $d$ is not a metric and it fails to be upper semicontinuous in general, while it is lower semicontinuous where it is finite. There is, however, a class of spacetimes where $d$ is finite and continuous, namely the \emph{globally hyperbolic} ones. These are defined to be causal\footnote{Originally, global hyperbolicity was defined using the stronger property of strong causality, which, however, has been shown to be equivalently replaceable by weaker conditions \cite{BS:07,Min:09a}, even weaker than causality \cite{HM:19}.} (i.e., there are no closed causal curves) with the so-called \emph{causal diamonds} $J(p,q):=J^+(p)\cap J^{-}(q)$ being  compact for all $p,q$. Also, by the \emph{Avez-Seifert theorem}, globally hyperbolic spacetimes are \emph{causally geodesically connected}, that is, any pair of points $p\leq q$ can be joined by a causal geodesic $\gamma$ that is maximising, i.e., $L(\gamma)=d(p,q)$. Conversely, and just as for minimising curves in the Riemannian case, maximising curves may be reparametrised as causal geodesics. 

The notion of \emph{geodesic completeness} in the Lorentzian case can be defined individually for spacelike, timelike, and null geodesics, giving rise to three independent notions. Also, causal geodesic completenss does not imply causal geodesic connectedness. So, in some sense, global hyperbolicity is the Lorentzian counterpart of Riemannian completeness, while \emph{incompleteness} is a central notion in the singularity theorems, as we shall discuss next. 

\subsection{Singularites and incompleteness}\label{sec:2.3}

Despite vivid examples such as the Schwarzschild metric and cosmological models of expanding universes, it is a somewhat subtle matter to define the general notion of a ``singularity'' in GR. Intuitively it should be a point where the curvature becomes unbounded but such a scenario is literally incompatible with the dynamical picture of spacetime in GR: We can only ever speak of a point in spacetime, if we have solved the field equations around it and hence found the corresponding manifold and metric structure. So, rather than being a point \emph{in} spacetime a singularity should be seen as some kind of singular boundary point of spacetime\footnote{Note that due to the failure of the Lorentzian distance to define a metric, a Cauchy completion is not available---another sharp contrast to the Riemannian world.}. The quest then is, how to detect the occurrence of a singularity from within spacetime, that is by properties of the spacetime itself---for a detailed discussion see e.g. \cite[Sec.\ 8.1]{HE:73} and the classic \cite{G:68}.
\medskip

To cut a long story (see e.g.\ \cite[Sec.\ 2.3]{Lan:21}) short, after Penrose's seminal paper \cite{Pen:65} it has---mainly under the influence of Hawking---become standard to define singularities via causal geodesic incompleteness. More precisely, we call a spacetime \emph{singular}, if it contains an incomplete causal geodesic.

In addition to giving a clear geometric condition, this definition is also physically reasonable: A future incomplete causal geodesic corresponds to a freely falling observer or to a light ray that suddenly ends its existence. In the past case it corresponds to an observer or light ray that suddenly pops into existence from nowhere. Both situations have to be considered as being even more objectionable than a blow up of curvature, and the general point of view has become to regard causal geodesic completeness as a minimal condition for a spacetime to be ``free of singularities''.

The above definition, however, also has severe drawbacks, see e.g. \cite[Sec.\ 3]{Sen:98} for a detailed discussion. In particular, the link between singularities and divergence of the curvature is lost. In fact, incompleteness could occur for trivial reasons, e.g.\ cutting out one point from an otherwise perfectly fine spacetime. This immediately leads us to the following notion: We call a spacetime \emph{extendible} if it can be isometrically embedded as an open submanifold into a larger spacetime\footnote{Sometimes also subject to specific regularity conditions, see also Section \ref{sec:4.1}, below.}. It is a fact, \cite[Prop.\ 6.16]{BEE:96} that timelike (or null) geodesic completeness implies inextendability. So we see that also in general, extendability can be a source of incompleteness and it becomes an important issue in the wake of the singularity theorems to exclude it. We shall return to this discussion in Section \ref{sec:4.1}, but we now head on to discuss the classical theorems.

\section{The classical singularity theorems}\label{sec:thms}

In this section we review the classical theorems. We start with the one of Hawking, which corresponds to the cosmological situation, since it is technically somewhat easier to formulate and to prove. We then proceed to the Penrose theorem which covers the gravitational collapse scenario, and finally, cover the most refined of the classical results, the Hawking-Penrose theorem. We start by revealing the general structure behind essentially all singularity theorems.

\subsection{The basic structure of the singularity theorems}

It has long been observed that all classical singularity theorems share the same structure. This point has been made most clearly by Jos\'e Senovilla, who formulated a ``pattern singularity theorem'' in \cite[Thm.\ 6.1]{Sen:98} to analyse the various statements, their conditions and their conclusions in a well organised way.

\begin{theorem}[Pattern singularity theorem] Let $(M,g)$ be a spacetime such that the following hold:
\begin{enumerate}
 \item [(E)] a condition on the curvatue, also called energy condition,
 \item [(C)] a condition on the causality, and
 \item [(I)] an initial or boundary condition.
\end{enumerate}
Then $M$ is causal geodesically incomplete.
\end{theorem}

Without specifying these conditions in detail we can nevertheless give a first idea of how the corresponding proofs work, that is, we explain how the conditions (E), (C), and (I) conspire to contradict geodesic completeness.

The initial condition (I) serves the purpose that some causal geodesics start to focus towards each other. 
%As an example we consider a spacelike hypersurface $\Sigma$ in spacetime with past pointing mean curvature vector $H$. Then future convergence
The energy condition (E) then implies that this focusing goes on until eventually a causal geodesic develops a focal or conjugate point and consequently stops maximising the Lorentzian distance. 
%Therefore we have found a geodesic that stops being maximizing. 
On the other hand the causality condition (C) implies the existence of maximising geodesics, at least in some region of spacetime. Now to dissolve this contradiction we have to accept that some geodesics stop existing before they reach a conjugate or focal point, that is, they are incomplete.
\medskip

Let us now examine each of the three conditions in some more detail and introduce the corresponding notions in a meaningful and precise manner. 

\subsection{Energy conditions and the focusing of geodesics}\label{sec:3.2}

We start with condition (E) which actually is a condition on the curvature of spacetime. It is, however, called energy condition since via the field equations \eqref{eq:ees} it corresponds to a condition on the energy-momentum tensor $T$. Physically these conditions roughly amount to the fact that gravity is attractive, at least on average,
and most ``reasonable'' matter models will satisfy them---for an extensive discussion see e.g.\ \cite[Sec.\ 4.3]{HE:73} and, in the context of the singularity theorems, \cite[Sec.\ 6.2]{Sen:98}. Here we only introduce the most important of these conditions. 

We say that a spacetime statisfies the \emph{strong energy condition (SEC)}, if 
\begin{equation}\label{eq:sec}\tag{SEC}
 \Ric(X,X)\geq 0\quad\mbox{for all timelike vectors $X$},
\end{equation}
and we say it satisfies the \emph{null energy condition (NEC)} if
\begin{equation}\label{eq:nec}\tag{NEC}
 \Ric(X,X)\geq 0\quad\mbox{for all null vectors $X$}.
\end{equation}
Note that by continuity \eqref{eq:sec} implies \eqref{eq:nec} and hence nonnegativity of $\Ric(X,X)$ for all causal $X$. So this terminology is consistent with the habit to formulate the \eqref{eq:sec} for all causal vectors\footnote{See, e.g. \cite[p.\ 95]{HE:73}, where ``our'' \eqref{eq:sec} is called the timelike convergence condition.}.

It is worthwhile to note that the energy conditions are the only place where the field equations come into play, so that the validity of the singularity theorems widely exceeds GR and continue to hold for alternative theories, as long as they imply the corresponding curvature conditions. 
\medskip

In the context of the singularity theorems the curvature condition is needed to exert a focusing effect on geodesics and we shall explain this in some detail, for more information see e.g.\ \cite[Sec.\ 4.6]{Kri:99}, \cite[Ch.\ 9]{BEE:96}. To begin with, we introduce the central notion of \emph{Jacobi fields}, i.e., vector fields $J$ along a geodesic $\gamma$ that satisfy the Jacobi equation 
\begin{equation}
 \ddot J + R(J,\dot\gamma)\dot\gamma=0\,,
\end{equation}
 where $\ddot{J}$ denotes the (iterated induced) covariant derivative along $\gamma$. Jacobi fields are in a one-to-one correspondence with \emph{geodesic variations} of $\gamma$, which picture the situation more vividly. Let $\gamma:[a,b]\to M$ be any curve, then a variation of $\gamma$ is a two-parameter map (with $\delta$ some small positive number)
 \begin{equation}\label{eq:variation}
  \mathbf{x}: [a,b]\times(-\delta,\delta) \to M
 \end{equation}
such that $\mathbf{x}(t,0)=\gamma(t)$ for all $a\leq t\leq b$, which we also call the base curve of $\mathbf{x}$. For fixed $s$ we call the $t$-parameter curves $t\mapsto \mathbf{x}(t,s)$ \emph{longitudinal}, and the $s$-parameter curves for fixed $t$, $s\mapsto \mathbf{x}(t,s)$ \emph{transverse}. The tangents of the transverse curves at the base curve give the \emph{variational vector field} $V(t)=\frac{d}{ds}\mid_{s=0}\mathbf{x}(t,s)$ along $\gamma$. Now given a geodesic $\gamma$ then any Jacobi field is the variational vector field of a geodesic variation of $\gamma$, that is, a variation where all longitudinal curves are geodesics as well. 

Essential for the focusing of geodesics is the notion of \emph{conjugate points}: two points $\gamma(a)$, $\gamma(b)$ are called conjugate along $\gamma$ if there exists a nontrivial Jacobi field vanishing at $a$ and $b$. This condition is equivalent to the existence of a geodesic variation $\mathbf{x}$ with $\mathbf{x}(a,s)=\gamma(a)$ for all $s$ and $V(b)=0$, which is further equivalent to the exponential map $\exp_{\gamma(a)}$ being singular at $b\,\dot\gamma(a)$. 

A geodesic variation as above with $V(b)=0$ can be pictured as an ``almost meeting point'' of nearby geodesics,  all starting out at $\gamma(0)$. Suppose for the moment that a meeting point actually occurs along a maximising causal geodesic $\gamma$, so that there is another maximising causal geodesic $\sigma$
%$\sigma=\mathbf{x}(.,\eta)$ 
with the same endpoints (i.e., $\sigma(a)=\gamma(a)$, $\sigma(b)=\gamma(b)$). Then $\gamma$ fails to maximise the Lorentzian distance beyond $\gamma(b)$ since the concatenation of $\sigma\mid_{[a,b]}$ with $\gamma\mid_{[b,b+\eta)}$ is a geodesic from $\gamma(0)$ to some $\gamma(b+\eta)$ with a break point at $\gamma(b)$ and the same length as $\gamma$. However, being broken it cannot be maximising and so $\gamma$ can't be either. Making this argument precise one arrives at the following central statement:

\begin{proposition}
 A causal geodesic fails to maximise the Lorentzian distance after its first conjugate point. 
\end{proposition}

From here it is evident that the curvature can be linked to the global structure of a spacetime by means of conjugate points. In fact, in the Riemannian case the analogous observation leads e.g.\ to the theorems of Hadamard and Myers, while in the Lorentzian case it is intimately related to the singularity theorems, cf.\ e.g.\ \cite{MS:15}. 

At this point it should perhaps be pointed out that a causal geodesic from a point stops being maximising if and only if either (a) there exists a distinct causal geodesic between the same endpoints of the same length (as argued above) or (b) the geodesic encounters a conjugate point.
%\footnote{A similar statement holds for causal geodesics emanating from a submanifold of $M$.}Given suitable geometrical conditions on the Lorentzian metric (e.g.\ a Ricci curvature bound, a ``convergence condition'' such as the existence of a trapped surface, and a completeness condition), one can use Riccati comparison techniques to show that all causal geodesics of a suitable type will encounter conjugate points, and hence stop being maximising curves between their endpoints. 
But the cut-locus of a point $p$ in a Lorentzian manifold (i.e., the set of points where the geodesics emanating from $p$ stop maximising) is a closed set of measure zero with the set of points which can be reached from $p$ by two distinct maximising geodesics being dense. Therefore, almost all geodesics that stop maximising do so due to (a) and hence do so 
%their intersection with another geodesic with the same endpoint of the same length. As such, 
%most causal geodesics will no longer be maximising 
even \emph{before\/} they encounter their first conjugate point. However, since (a)
%such an intersection of geodesics 
is related to the~\emph{global\/} geometry of the manifold, there is no way to detect such points via estimates on the curvature. So
%estimate (in terms of, say, the curvature) the distance that one must traverse along a given curve before one encounters such an intersection. 
the power of conjugate points lies the fact that they lead to geodesics no longer being maximising \emph{and\/} that we can estimate when they occur.
% \todo{Indeed one may say that there are three reasons why focal points may occur ???? find formulation or just say that there are no ic's}.
\medskip

Motivated by this observation we now introduce the analytical tools to find conjugate points, for a detailed and systematic analysis based upon Lorentzian Morse index theory see \cite[Ch.\ 10]{BEE:96}, \cite{Oha:22}.

To begin with, we observe that the relevant information on conjugate points is contained in the $(n-1)$-dimensional subspace of Jacobi fields vanishing at a given point and taking values in $\dot\gamma(t)^\perp:=\{v\in T_{\gamma(t)}M:\, \langle v,\dot\gamma(t)\rangle=0\}$. Now, to also properly deal with the null case (where $\dot\gamma(t)\in\dot\gamma(t)^\perp$) it is useful to work on the quotients
\begin{equation}
 [\dot\gamma(t)]^\perp:=\dot\gamma(t)^\perp/{\mathbb R}\dot\gamma(t)\quad\mbox{and}\quad 
 [\dot\gamma]^\perp:=\cup_t[\dot\gamma(t)]^\perp,
\end{equation}
respectively. Note that in the timelike case $[\dot\gamma(t)]^\perp$ coincides with $\dot\gamma(t)^\perp$ hence is of dimension $n-1$, while in the null case its dimension is $n-2$. We will henceforth use the letter $d$ to denote the dimension of $ [\dot\gamma]^\perp$ simultaneously in both cases. Observe that the restriction of the metric $g\mid_{[\dot\gamma]^\perp}$ is positive definite in both cases.

Next, to represent the information contained in Jacobi fields most economically we collect them into Jacobi tensor classes. That is, classes of $(1,1)$-tensor fields $[A]:[\dot\gamma]^\perp\to [\dot\gamma]^\perp$ for which the tensor Jacobi equation 
\begin{equation}
 [\ddot A]+[R][A]=0,\ \mbox{with}\  [R]:[v]\mapsto[R(v,\dot\gamma)\dot\gamma]\ \mbox{the \emph{tidal force operator}}
\end{equation}
holds and that in addition satisfy the nontriviality condition $\ker [A(t)]\cap\ker[\dot A(t)]=\{0\}$ for all (or equivalently just one) $t$. A Jacobi tensor (class) can be viewed as a matrix with its columns given by (classes of) Jacobi fields. In particular, given $[Y]\in[\dot\gamma]^\perp$ parallel (i.e., $[\dot Y]=0$), then $[A]([Y])$ is a Jacobi field. Moreover, by the nontriviality condition it is nontrivial provided $Y\not=0$. Therefore $\gamma(a)$ is conjugate to $\gamma(b)$ iff the (unique) Jacobi tensor class with $[A(a)]=0$ and $[\dot A(a)]=\id$ satisfies $\ker[A(b)]\not=\{0\}$.

Given a Jacobi tensor class $[A]$, then $[B]:=[\dot A][A^{-1}]$ satisfies the \emph{matrix Riccati equation}
\begin{equation}\label{eq:riccati}
 [\dot B]+[B]^2+[R]=0.
\end{equation}
A Jacobi tensor class is called \emph{Lagrange} if its \emph{Wronskian} vanishes, i.e., $W([A],[A]):=[\dot A^{\dagger}][A]-[A^{\dagger}][\dot A]=0$ (with $A^\dagger$ denoting the adjoint), and this is the case if $[A(t)]=0$ for some $t$. As a consequence $B$ (wherever defined) is then self adjoint and the analytic way to detect conjugate points is encoded in the vorticity-free \emph{Raychaudhuri} equation for the \emph{expansion} $\theta:=\tr([B])=\tr([\dot A][A]^{-1})=(\det [A])^{-1}\,(\det [A])\dot{}$ 
%of a Jacobi tensor class, i.e.,
\begin{equation}\label{eq:ray}
 \dot\theta=-\Ric(\dot\gamma,\dot\gamma)-\tr(\sigma^2)-\frac{\theta^2}{d},
\end{equation}
where the shear $\sigma$ is defined as $\sigma:=1/2([B]+[B^\dagger])-(\theta/d)\id$. 

The key observation now is the following: The second and the third term on the right hand side of \eqref{eq:ray} are non-positive. If we assume the first term to be non-positive as well by demanding that $\Ric(\dot\gamma,\dot\gamma)\geq 0$, e.g.\ via the energy conditions \eqref{eq:sec} or \eqref{eq:nec}, we can easily ``generate'' conjugate points. Indeed, if $\theta(a)$ is negative at some parameter value $a$, then it will diverge to $-\infty$ in finite parameter time. More precisely, \eqref{eq:sec}/\eqref{eq:nec} and \eqref{eq:ray} imply $\dot\theta\leq -\theta^2/d$ which upon integration from $a$ to some $t>a$ gives
\begin{equation}\label{eq:ray-est}
 \theta\leq \frac{d}{t-a+d/\theta(a)}
\end{equation}
and so $\theta$ diverges for some $t\in[a,a-d/\theta(a))$. 
Consequently if $[A]$ is a Jacobi tensor class with $[A](a)=0$ and $[\dot A](a)=\id$ and we have $\mid\theta(t)\mid\to\infty$ for $t$ to some $b$, then $\det [A(b)]=0$ 
%and $[A(b)]$ has a nontrivial kernel 
and so $\gamma(b)$ is conjugate to $\gamma(a)$. Hence we have established:

%BEE 12.9
\begin{proposition}[Conjugate points from negative expansion]\label{prop:ecp}
 Let $\gamma:I\to M$ be an inextendible causal geodesic and suppose there is a Lagrange tensor class $[A]$ with negative expansion $\theta(a)=\tr([\dot A(a)][A(a)]^{-1})$ at some $a\in I$. If $\Ric(\dot\gamma(t),\dot\gamma(t))\geq 0$ for all $t$ then there is a point conjugate to $\gamma(a)$ along $\gamma$ at some $b<a-d/\theta(a)$, provided $b\in I$.
\end{proposition}

\medskip

To sum up, we have seen that if along a causal geodesic a negative expansion occurs, then given the energy conditions, eventually a conjugate point will arise implying the failure of the geodesic to be maximising beyond it. Now, to ``generate'' a point with negative expansion is the purpose of the initial conditions to which we turn next.

\subsection{Initial conditions}

We now proceed to condition (I), which is the ingredient that gets the focusing effect of geodesics started, i.e., which produces a negative $\theta(a)$ at some point $\gamma(a)$ along a causal geodesic $\gamma$. 

Beginning  with the simplest case, consider a spacelike hypersurface $S$ in spacetime (i.e., $g\mid_{TS}$ Riemannian) and its future directed unit normal $\nu$, i.e., $\langle\nu,\nu\rangle=-1$. Now at a given point $p$ in $S$ we consider the geodesic $\gamma_\nu$ starting in $p$ with future normal unit velocity, i.e., $\gamma_\nu(0)=p$, $\dot\gamma_\nu(0)=\nu(p)$. Further consider a Jacobi tensor\footnote{Since we are dealing with a timelike geodesics here, there is no need to consider tensor classes.} $A$ along $\gamma_\nu$ with $A(0)=\id$ and $\dot A(0)=-S_\nu(p)$, where $S_\nu$ is the shape operator of $S$ given by $S_\nu(X):=-\nabla_X\nu$. A standard calculation now gives 
%\todo{check signs once more; cf. O'Neill, p.431}
\begin{equation}\label{eq:theta}
 \theta(0)=\tr(-S_\nu)\lvert_p=-(n-1)\langle H(p),\nu(p)\rangle=:-(n-1)k(p),
\end{equation}
where $H$ is the \emph{mean curvature vector} of $S$ and we call $k$ the future \emph{convergence} of $S$. 

Now, if we suppose that $H(p)$ is past pointing timelike or, equivalently, that the convergence $k(p)$ is positive, we have achieved $\theta(0)<0$ and we have kicked off focusing. 

To formulate a corresponding statement we still have to introduce the analogue of a conjugate point along a geodesic in the current ``endmanifold'' case, i.e., the concept of a \emph{focal point}. More precisely, a point $q$ is called focal to a hypersurface $S$ along a timelike geodesic $\gamma$ that starts at some $p\in S$ and normal to it, if there is a nontrivial normal Jacobi field $J$ along $\gamma$ with $J(0)=0$ and $J'(0)=\nabla_{J(0)}\nu$ that vanishes at $q=\gamma(b)$. Analogously to the case of a conjugate point, such a Jacobi field corresponds to the variational vector field $V$ of a variation $\mathbf{x}$ of $\gamma$ through geodesics that all start at $S$ and normal to it with $V(b)=0$. Equivalently the normal exponential map $\exp_p^\perp:S_p^\perp\to M$ that takes a normal vector $v\in S_p^\perp\subseteq T_pM$ to $\gamma_v(1)$, is singular at $b\,\dot\gamma(0)$. Now, just as in the conjugate-point case, a geodesic $\gamma_\nu$ starting normal to a hypersurface $S$ stops maximising the Lorentzian distance to $S$, denoted by $d(S,.)$, after its first focal point. Also observe that %(just as in the Riemannian case) 
if a geodesic does not start out normally from $S$, it is not maximising at all.
A typical focusing result then reads as follows:

\begin{proposition}[Focal points from negative expansion]\label{prop:eofp}
Let $\gamma:[0,\beta)\to M$ be an inextendible timelike geodesic starting at $p$ normally from a spacelike hypersurface $S$ and suppose $\Ric(\dot\gamma,\dot\gamma)\geq 0$ for all $t$. If $\theta(0)$ is negative (equivalently if the convergence $k(p)$ is positive) then there is a focal point $\gamma(b)$ to $S$ for some $b\leq-(n-1)/\theta(0)$, provided $b<\beta$. Consequently, $\gamma$ stops maximising the Lorentzian distance to $S$ (the latest) at $b$ if it exists that long.
\end{proposition}

The above focusing result will turn out to be at the analytical core of Hawking's singularity theorem, and we now turn to the initial condition for Penrose's theorem. In fact, the notion of a \emph{trapped surface}---to be introduced now---is one of the great innovations of the celebrated paper \cite{Pen:65} and its influence on the development of GR cannot be overestimated, see e.g.\ \cite[Sec.\ 7.2]{SG:15} for a brief description of its impact. It gives a precise mathematical formulation to the idea of a spacetime region of so strong gravity that not even light can escape the gravitational pull. Its key model are the ``trapped spheres'' inside the horizon of a Schwarzschild black hole.

To begin with, we consider a spacelike submanifold $P$ of codimension $2$.
Since its normal bundle in $M$ is Lorentzian, there are at any point of $P$ two linearly independent future pointing null vectors  normal to $P$. We call them $\nu_1$, and $\nu_2$ and normalise them via the condition $\langle \nu_1,\nu_2\rangle=-1$.
The corresponding null vector fields on $P$ give rise to two distinct families of null geodesics which, physically speaking, represent the two families of light rays emanating orthogonally from $S$ into the future. Interpreting $P$ for the moment as the surface of a star at a given instant of time, our experience leads us to think that the congruence of light rays sent towards the center of the star should converge, while the second ``outgoing'' congruence should diverge. However, if the effect of gravity is strong---which in the present picture manifests itself in the extrinsic curvature of the spacelike hypersurface representing the instant of time---both congruences will converge. In fact, this is the condition of $P$ being future trapped.
Mathematically we can expressed this by demanding the expansions of \emph{both} the families of normal future pointing null geodesics $\theta_1$ and $\theta_2$ to be negative on all of $P$. Equivalently,  we can demand the mean curvature vector of $S$ to be past pointing timelike.

This alone is not a useful property as is demonstrated by the trivial example of the intersection of two past lightcones in Minkowski space. However if we ask the hypersurface to be closed, i.e., compact without boundary (as suggested in the above example) it becomes a mighty tool as we shall see below. 

To get a vivid picture of the contraction property, we express it via the variation of the area $A$ of $P$ along the flow of a vector field $\xi$, which by a standard formula is given by %\todo{recheck sign!}
\begin{equation}
 \delta_\xi A=-\int_A \langle H,\xi\rangle.%, \quad\mbox{where $\eta=\langle\xi,\xi\rangle$}.
\end{equation}
This equation, in the Riemannian case, of course expresses the fact that minimal surfaces ($H=0$) are critical points of the area functional. In the Lorentzian situation, however, $H$ being past pointing and timelike implies that the variation of the area along any future directed null vector field $\xi$ is negative.
\medskip

We finally give the official definition of a closed trapped submanifold of arbitrary codimension $1<m<n$, and reserve the term closed trapped surface for the case $m=2$.

\begin{definition}[Closed trapped submanifold]
 A compact without boundary, spacelike submanifold $P$ is called \emph{closed future trapped submanifold} if 
 its mean curvature vector field is past pointing timelike on all of $P$.
 \end{definition}
 
% Again equivalent conditions of course are that for any normal future directed null vector field $\nu\in TP^\perp$ its convergence $k_s(\nu)=\langle H,\nu\rangle$ is positive, and that the null second fundamental form associated with any $\nu$ as above have negative trace, respectively.  
% \medskip

The idea of lightrays being trapped can also be expressed in purely causal terms via the notion of a future \emph{trapped set}, i.e., a closed, achronal set $A$ such that its \emph{future horismos} 
\begin{equation}
E^+(A):=J^+(A)\setminus I^+(A) 
\end{equation}
is compact. Here achronal means that no two points in $A$ are timelike related. By push up $E^+(A)$ is achronal, and if it is compact, then $A$ has to be compact itself since $A\subseteq E^+(A)$. Observe that in general $E^+(A) \subsetneq \partial J^+(A)=\partial I^+(A)$, although equality holds locally and also in globally hyperbolic spacetimes. 

Now, using a focusing argument much like the one that lead to Proposition \ref{prop:eofp} one may show that the analytic concept of a closed trapped surface $P$ under appropriate conditions implies the causal concept of trappedness. Indeed, by a variational argument $E^+(P)$ is generated by conjugate-free null geodesics: to any point $q\in E^+(P)$ there runs a null geodesic from $P$ in $E^+(P)$ to $q$ that has no conjugate points before $q$. Now by focusing all these null geodesics eventually do develop conjugate points which implies that $E^+(P)$ is contained in the normal exponential image of a compact set and by closedness is compact itself. We have hence argued for the following result to hold, which can be generalised to trapped submanifolds of codimension $m>1$ under an additional curvature condition, see \cite[Prop.\ 3]{GS:10}:
%Go back to the simpler version!
% More precisely, in a future null complete spacetime where \eqref{eq:nec} holds, any achronal closed future trapped surface is also a future trapped set. A similar statement additionally uses the notion of strong causality, which means that there are no almost closed causal curves in $(M,g)$ in  the sense that every point $p$ has arbitrarily small neighbourhoods that no causal curve intersects in a disconnected set. It may then be shown that $S=E^+(P)\cap P$ is a future trapped set for any closed future trapped surface $P$, that is, we have:
% 
% %O'neill 14.60 needs achronality of the trapped srfce which can be circumvented, see BEE, 12.36 and 12.45
% \begin{proposition}[Trapped set from trapped surface]\label{prop:tsts}
% Let $(M,g)$ be a strongly causal and future null complete spacetime where \eqref{eq:nec} holds. If $(M,g)$ contains a closed future trapped surface then it also contains a future trapped set.
% \end{proposition}
\begin{proposition}[Trapped set from trapped surface]\label{prop:tsts}
 Let $(M,g)$ be a future null complete spacetime where \eqref{eq:nec} holds. Then any achronal closed future trapped surface is also a future trapped set. 
\end{proposition}

The occurrence of geodesic focusing as such, however, does not lead to a singularity, and that is where some more causality theory comes into play.

\subsection{Causality theory}\label{sec:3.4}

In the discussion leading to Proposition \ref{prop:tsts} we have used the inexistence of conjugate 
points in $E^+(P)$. 
%current discussion causality conditions are needed 
More generally, we need to exclude the existence of focal or conjugate points in certain subdomains of spacetime to infer the existence of an incomplete geodesic from geodesic focusing. 

To give a proper account we first introduce the notion of a \emph{Cauchy surface}, which informally can be interpreted as an ``instance of time`` serving as ``initial surface'' when formulating the Einstein equations as an evolutionary system. Formally it is an achronal (i.e., no two points are chronologically related) closed (Lipschitz) hypersurface of spacetime that is hit by every inextendible causal curve. 
%It follows that any Cauchy surface $S$ is \emph{achronal}. 
A central fact is that a spacetime admits a Cauchy surface iff it is \emph{globally hyperbolic}, a statement which connects causal properties of spacetime with PDE-theory.

% that is if there are no closed timelike curves \todo{check} and the so-called causal diamonds $J(p,q):=J^+(p)\cap J^-(q)$ are compact. It is this compactness property which makes this notion an analog to Riemanninan completeness in the sense that it allows to ``save'' the geodesic connectedness-part of the Hopf-Rinow theorem. In fact, the Avez-Seifert theorem asserts that in a globally hyperbolic set any pair of causally related points can be joined by a maximising causal geodesic. 

Now global hyperbolicity is a strong condition and in many situations one wants certainly to do without it. Therefore the following generalisation comes in handy: We define the future \emph{Cauchy development} $D^+(A)$ of any achronal set $A$ as
\begin{equation}\nonumber
 D^+(A):=\{p\in M:\ \mbox{Every past inextendible causal curve through $p$ meets $A$}\}.
\end{equation}
We define the past Cauchy development $D^-(A)$ analogously, set $D(A):=D^+(A)\cup D^-(A)$ and call it the Cauchy development of $A$. It is then a result that the interior ${D}(A)^\circ$ of the Cauchy development is globally hyperbolic, generalising the situation of a Cauchy surface, whose Cauchy development is the entire, globally hyperbolic spacetime. More specifically, for any acausal (i.e., no two points are causally related) topological hypersurface $S$, its Cauchy development is open and globally hyperbolic. Now, compactness of the causal diamonds allows one to show:

\begin{proposition}[Existence of maximiser]\label{prop:eom}
 Let $S$ be a closed, achronal, spacelike hypersurface. Then to any point $p$ in the future Cauchy development $D^+(S)$ there runs a future directed maximising geodesic $\gamma$ from $S$. Moreover, $\gamma$ starts normal to $S$, has no focal point before $p$, and it is timelike unless $p\in S$.
\end{proposition}

Finally, we need to consider the boundary of the Cauchy development, called the \emph{Cauchy horizon}. More precisely, this notion is defined in causal terms as follows. The future Cauchy horizon of an achronal set $A$ is given by
\begin{eqnarray}\nonumber
 H^+(A)&=&\overline{D^+(A)}\setminus I^{-}(D^+(A))\\
 &=&\{p\in\overline{D^+(A)}: \ \mbox{$I^+(p)$ does not meet $D^+(A)$}\}.
\end{eqnarray}

It is now a basic fact that $H^+(S)=I^+(S)\cap\partial D^+(S)$, and with these preparations we are now ready to proceed from the pattern theorem to the ``real'' %singularity 
theorems.

% We collect some basic properties of Cauchy horizons.
% \begin{lemma}[Properties of Cauchy horizons]\label{lem:poch}
%  For a (topologically) closed, acausal $C^0$-hypersurface we have
%  \begin{enumerate}
%   \item
%   \item $H^+(S)$, if nonempty, is a closed achronal topological hypersurface.
%   \item At each point $p\in H^+(S)$ there is a past inextendible null geodesic without conjugate points that is entirely contained in $H^+(S)$ (a so-called null generator).
%  \end{enumerate}
% \todo{all needed?}
% \end{lemma}

%\todo{what else needed?}
% An acausal edgeless (and hence closed) set is a partial Cauchy hyper-
% surface.
%\medskip

\subsection{The three classical theorems}\label{sec:3.5}

The first one of the classical statements we wish to discuss is the one by Hawking, which technically is the easiest, and we will be able to provide a sketch of its proof along the above discussion. In the literature 
one actually finds several (versions of) singularity theorems associated with the name of Hawking and we wish to discuss the one that supposes the existence of a compact Cauchy surface. It is, in particular, applicable to a spatially closed universe and hence gives evidence for a big bang in such models. It has, however, become a custom\footnote{At least in the more mathematically oriented literature.} to formulate it in a time-reversed manner, predicting a future singularity.

\begin{theorem}[Hawking]\label{thm:CH} 
Let $(M,g)$ be a spacetime such that 
\begin{itemize}
 \item [(E)]  $\mathrm{Ric}(X,X)\geq0$ for all timelike vectors $X$, i.e., \eqref{eq:sec} holds, 
 \item [(C)]  there is a compact spacelike Cauchy surface $S$ in $M$, with
 \item [(I)]  everywhere positive future convergence $k$.
\end{itemize}
Then $M$ is future timelike geodesically incomplete.
\end{theorem}

\noindent
\emph{Sketch of proof.} By compactness there is a positive minimum $k_0$ of $k$ on $S$. So by Proposition \ref{prop:eofp} every timelike geodesic starting normally from $S$ encounters a focal point at $t=1/k_0$ the latest. On the other hand by Proposition \ref{prop:eom} every $p\in D^+(S)\setminus S$ is reached by a normal timelike geodesic without focal point and therefore
\begin{equation}
 D^+(S)\subseteq\{p\in M:\ d(S,p)\leq 1/k_0\}.
\end{equation}
If a future directed timelike curve starting in $S$ were to leave $D^+(S)$ it had to pass through $\partial D^+(S)$ and hence %by Lemma \ref{lem:poch}(i) 
the Cauchy horizon $H^+(S)$ would be non-empty. However, being a Cauchy surface, $S$ has empty Cauchy horizon and so $I^+(S)\subseteq D^+(S)$. This, in particular, forces the above geodesics to have finite length,  which makes them incomplete.\hfill \qed
\medskip

Next we turn to the theorem of Penrose. As already indicated above, its initial condition is the existence of a trapped surface, but
%which is physically taken to be the indication that gravitational collapse is underway \todo{adapt}. T
the theorem has been generalised to \emph{trapped submanifolds} of arbitrary codimension in \cite[Thm.\ 1]{GS:10}. We will, however, include this advancement explicitly only in in our statement of the Hawking-Penrose theorem.  

At the heart of the proof of the Penrose theorem lies the following fact: The topological condition (existence of a trapped set) that is derived from the existence of a trapped surface, contradicts the existence of a non-compact Cauchy surface. The latter condition physically amounts to the fact that we are considering an isolated system and, in particular, the gravitational collapse of an isolated body.

\begin{theorem}[Penrose]\label{thm:CP} 
Let $(M,g)$ be a spacetime such that
 \begin{itemize}
 \item [(E)]  $\mathrm{Ric}(X,X)\geq0$ for all null vectors $X$, i.e., \eqref{eq:nec} holds, 
 \item [(C)]  there is a non-compact Cauchy surface $S$, and
 \item [(I)]  there is an achronal closed future trapped surface $P$. 
\end{itemize}
Then $M$ is future null geodesically incomplete.
\end{theorem}

\noindent
\emph{Sketch of proof.} Indirectly assuming completeness, we first establish that $E^+(P)$ is a non-empty, compact, and achronal topological hypersurface: By Proposition \ref{prop:tsts} $E^+(P)$ is compact and global hyperbolicity implies that $E^+(P)=\partial J^+(P)$. The latter is nonempty (by achronality of $P$) and being the boundary of a so called future-set (i.e., a set containing its own chronological future) it is a $C^0$-hypersurface (see also below).

Now we take any timelike vector field $Y$ on $M$ (whose existence is guaranteed by time-orientability) and define the map $\rho:\, M\to S$, taking each $p$ to the unique intersection point of the maximal integral curve of $Y$ though $p$ with $S$. Obviously $\rho$ leaves $S$ invariant, and one may show that it is continuous and open. The restriction of $\rho$ to $\partial J^+(P)$ is injective by achronality and open between topological hypersurfaces and so by invariance of domain $\rho(\partial J^+(P))$ is open. On the other hand by compactness it is also closed and so $\rho(\partial J^+(P))=S$. But since the latter set is non-compact, we have reached a contradiction.\hfill\qed
\medskip

For the final part of this section we turn to the most sophisticated of the classical singularity theorems, namely to the one by Hawking and Penrose \cite{HP:70}. It collects the various developments and variants of singularity theorems that have appeared in the years before and recovers most of them under much weaker assumptions. In particular, the assumption of global hyperbolicity is avoided throughout. Consequently the necessary focusing results require a deeper analysis of the influences of curvature on causal geodesics and we will discuss this issue in some detail. In particular, a new assumption called \emph{genericity condition} is introduced, which guarantees that all causal geodesics $\gamma$, at least at one point $\gamma(t)$ ``feel'' the effect of curvature in the sense that at $\gamma(t)$ the tidal force operator is nontrivial,  for more details see \cite[Sec.\ 2.5]{BEE:96}.
\medskip

Also the Hawking-Penrose theorem features a third possible initial condition: a (future) trapped point\footnote{For its physical significance see e.g.\ \cite[p.266]{HE:73}.}, i.e., a point $p$ such that the expansion becomes negative for any (future directed) null geodesic starting in $p$. Also the initial condition of the Hawking theorem is generalised from the existence of a compact Cauchy surface to merely a compact partial Cauchy surface, i.e., an \emph{achronal set without edge}. Here the edge of an achronal set $A$ consists of all points $p\in\overline A$ possessing arbitrarily close pairs of points $x\in I^-(p)$, $y\in I^+(p)$ that can be connected by a timelike curve not intersecting $A$. An achronal set can be shown to be a closed topological hypersurface iff its edge is empty. This also implies that the boundary of a future set is a closed achronal topological hypersurface---a result already used in the proof of the Penrose theorem. 

Later Dennis Gannon \cite{Gan:75} and Charles Walter\footnote{In an earlier version of this paper 
%---following the AMS database at \url{http://www.ams.org/mathscinet}---
I misattributed this work to Chong Wan Lee. I wish to thank the true author to point me at this glitch.}  Lee \cite{Lee:76} independently established results under yet another kind of initial condition: the occurrence of some nontrivial topology in a compact region of a (partial) Cauchy surface. Hence this body of results---often termed the Gannon-Lee theorems---establish a firm link between the topological and the singularity structure of spacetime.
\medskip

Another more technical innovation of the Hawking-Penrose ``singularity theorem par excellence''\footnote{In the words \cite[p.\ 790]{Sen:98}.} is that the causal part of the argument was outsourced to the following separate statement\footnote{In \cite[p.\ 538]{HP:70} this statement is called the theorem while the actual Hawking-Penrose theorem appears as a corollary on p.\ 544.}: 

\begin{lemma}[Hawking \& Penrose]\label{lem:hp}%\cite[pp.~538]{HP}
In any spacetime $(M,g)$ the following three statements cannot simultaneously hold:
\begin{enumerate}
\item[(C1)] \label{lem:HP1} $M$ is chronological, i.e., it contains no closed timelike curve.
\item[(C2)]\label{lem:HP2} Every inextendible causal geodesic in $M$ contains a pair of 
conjugate points.
\item[(C3)] \label{lem:HP3} There is trapped set $A$.
\end{enumerate}
\end{lemma}

This Lemma goes well beyond the causality arguments presented in this review so far and we will not attempt to sketch a proof here. Note that the causality condition (C1) is way more general than global hyperbolicity which was used in the above theorems. Nevertheless, it can be generalised somewhat further, see the discussion in \cite[p.\ 793]{Sen:98}.

We now give the analytic result in an extended form which, as already announced, includes yet another class of initial conditions, namely trapped submanifolds of co-dimension $2<m<n$, which is due to \cite[Thm.\ 3]{GS:10}\footnote{Since the new condition \eqref{eq:gs} is actually redundant in the classical cases $m\in\{1,2,n\}$ \cite[Rem.\ below Thm.\ 3]{GS:10} we could have omitted to state the latter at all, but have chosen not to do so for the sake of presentation.}.   

\begin{theorem}[Hawking \& Penrose]\label{thm:CHP}
Let $(M,g)$ be a spacetime such that
 \begin{itemize}
 \item[(E)] the strong energy condition \eqref{eq:sec} holds
 %, i.e., $\mathrm{Ric}(X,X)\geq 0$ for all timelike vector fields $X$,
 %\item[(E2)] 
 as well as the genericity condition along any causal geodesic $\gamma$, i.e., there is a point $\gamma(t_0)$ such that 
 \begin{equation}\label{eq:gc}\tag{GC}
   [R (.,\dot{\gamma})\dot{\gamma}\mid_{t_0}]:\, [\dot\gamma(t_0)]^\perp\to [\dot\gamma(t_0)]^\perp
   \quad\mbox{is nontrivial, and}
 \end{equation}
 % \footnote{Here we have used components w.r.t.\ an arbitrary basis and the square bracket denotes antisymmetrisation.}
%\begin{equation}
%\dot{\gamma}^c \dot{\gamma}^d \dot{\gamma}_{\left[a\right.}R_{\left. b \right]cd\left[e\right.}\dot{\gamma}_{\left. f %\right]} \neq 0,
%\end{equation}
\item[(C)] it is chronological.
\end{itemize}
Moreover, assume it contains at least one of the following:
\begin{enumerate}
\item[(I1)] a compact achronal set without edge,
\item[(I2)] a closed future trapped surface $P$,
\item[(I3)] a closed future trapped submanifold $P$ of co-dimension $2 < m < n$ such that additionally
 \begin{equation}\label{eq:gs}
  \sum_{i=1}^{n-m} \langle R(E_i,\dot{\gamma})\dot{\gamma},E_i \rangle
  \geq 0
 \end{equation}
 for any future directed null geodesic with $\dot{\gamma}(0)$ orthogonal to $P$, or
\item[(I4)] a future trapped point, i.e., $p\in M$ such that on every future directed null geodesic from $p$ the expansion $\theta $ becomes negative.
\end{enumerate}
Then $M$ is causal geodesically incomplete.
\end{theorem}

Observe that we have not stated \emph{future} causal incompleteness, which, however, holds in the respective future cases of the assumptions, see \cite[p.\ 792, bottom]{Sen:98} for details.
\medskip

The line of arguments proving Theorem \ref{thm:CHP} from Lemma \ref{lem:hp} then is as follows: Obviously the causality condition (C) and (C1) agree. The genericity condition~\eqref{eq:gc} and \eqref{eq:sec} are used to show that: 
\begin{enumerate}
 \item [(1)] every inextendible causal geodesic in $M$ contains a pair of conjugate points, i.e., (C2) holds, and
 \item [(2)] any of the initial conditions (I1)-(I4) imply that there is a trapped set, i.e., that (C3) holds.
\end{enumerate}

Let us elaborate somewhat on these two items. Starting with (1), the trick is done via the following advanced focusing result:

\begin{proposition}[Focal points from genericity]\label{prop:ecp2}
 Let $\gamma$ be a complete causal geodesic and assume \eqref{eq:gc} and $\Ric(\dot\gamma(t),\dot\gamma(t))\geq 0$ for all $t$. Then $\gamma$ has a pair of conjugate points.
\end{proposition}

Observe that this result is much stronger than Proposition \ref{prop:ecp} since we do not assume the existence of a point with negative expansion. This is actually why a more detailed analysis of the influence of the curvature on causal geodesics is needed and the full matrix Riccati equation \eqref{eq:riccati} has to be used rather than just its trace, the (vorticity free) Raychaudhuri equation \eqref{eq:ray}. The key step in the proof, which is long and technical  (cf.\ e.g.\ \cite[p.\ 436--443]{BEE:96}), is to establish that along any geodesic as in the statement with the tidal force operator nontrivial at some $\gamma(t_0)$ (which exists due to \eqref{eq:gc}) we have
\begin{align}\nonumber
  &\mbox{All Lagrange tensor classes $[A]$ along $\gamma$}\\ \label{eq:lpm}
  &\mbox{with $[A(t_0)]=\id$ and $\theta (t_0)\leq 0$ become singular for some $t>t_0$ }
\end{align}
(and analogously for $\theta(t)\geq0$ and $t<t_0$).\medskip

\noindent
\emph{Proof of \eqref{eq:lpm}.}
First note that \eqref{eq:ray} and \eqref{eq:sec} give $\dot\theta(t)\leq 0$ for all $t$ and so $\theta(t)\leq 0$ for all $t\geq t_0$. If $\theta(t)<0$ for some $t\geq t_0$ then Proposition \ref{prop:ecp} gives the claim. 

So assume that $\theta(t)=0$ for all $t\geq t_0$ implying that also $\dot\theta(t)$ vanishes there. Inserting again into \eqref{eq:ray} and using \eqref{eq:sec} we obtain $0\leq -\tr(\sigma^2(t))$ for all $t\geq t_0$. But $\tr(\sigma^2(t))$ is non-negative, and so it has to vanish for $t\geq t_0$. Consequently for $t\geq t_0$ the shear $\sigma=1/2([B]+[B^\dagger])$ being self-adjoint, has to vanish itself and since $[B]$ is self-adjoint as well, it too has to vanish. But this implies by the Riccati equation \eqref{eq:riccati} that the tidal force operator is trivial at $t_0$, which contradicts our initial assumption.\hfill\qed 
\medskip

We now turn to the discussion of (2) and begin with the following extension of Proposition \ref{prop:tsts}:
%(cf.\ {BEE:96}, Prop.\ 12.36 (i), Prop.\ 12.46 for (iii) and \cite[Prop.\ 4]{GS:10} for (ii)):
\begin{proposition}[Comapact horismos]\label{prop:3.13}
 Suppose \eqref{eq:nec} and the existence of either (i) a closed future trapped surface $P$, or (ii) a closed future trapped submanifold $P$ as in (I3), or (iii) a future trapped point $P$. Then $E^+(P)$ is compact, or the spacetime is null incomplete.
\end{proposition}

% Alternative version dumped
% We now turn to the discussion of (2) and begin with the following extension of Proposition \ref{prop:tsts}, which, however, needs the notion of strong causality, which means that there are no almost closed causal curves in $(M,g)$ in the sense that every point $p$ has arbitrarily small neighbourhoods that no causal curve intersects in a disconnected set. It is then a fact that (C1) and (C2) imply strong causality and so using it in the following proposition does not harm us.
% %(cf.\ {BEE:96}, Prop.\ 12.36 (i), Prop.\ 12.46 for (iii) and \cite[Prop.\ 4]{GS:10} for (ii)):
% \begin{proposition}[Comapact horismos]
%  Suppose $(M,g)$ is a strongly causal and null complete spacetime where the \eqref{eq:nec} holds. If $M$ contains either (i) a trapped surface, or (ii) a trapped submanifold as in (I3), or (iii) a trapped point, then it also contains a trapped set.
% \end{proposition}

The proof uses again the focusing argument Proposition \ref{prop:ecp} to establish the main point, i.e., that $E^+(P)$ is contained in  the compact subset $\exp([0,T]K)$ for some compact $K$. 
%\todo{leaving $J^+$ to be explained somewhere...}

Now given (I4) condition (C3) follows immediately. In the cases (I2), (I3) one is only almost there, since in general $P$ need not be achronal. The trick here is to establish that $S=E^+(P)\cap P$ is achronal with compact horismos, hence a trapped set. This can be done 
%(\cite[Prop.\ 12.45]{BEE:96}, resp.\ \cite[Prop.\ 4]{GS:10}) 
under the additional assumption of strong causality\label{str-caus}, which, however in chronological spacetimes is a consequence of \eqref{eq:sec}, \eqref{eq:gc}, and null completeness. 
% see \cite[Prop.\ 12.40]{BEE:96}).     

Finally in case (I1) one shows that a compact achronal set $P$ without edge is a topological hypersurface with $E^+(P)=P$ which, again gives (C3). 
\medskip

This finishes our sketch of the arguments that establish the Hawking-Penrose theorem and also our journey into the classical singularity theorems. We next turn to their low-regularity extensions.

\section{Low regularity singularity theorems}\label{sec:lr}

In this section we wish to provide an overview of the extension of the classical singularity theorems to metrics of low regularity that have emerged over the last couple of years. Indeed the three key theorems discussed above have first been generalised to Lorentzian metrics of regularity $C^{1,1}$ \cite{KSSV:15,KSV:15,GGKS:18} and then in a further effort to $C^1$-metrics \cite{G:20,KOSS:22} with a Gannon-Lee theorem proved in \cite{SS:21}. We shall review the main mathematical advances that were developed to arrive at these results, mainly concentrating on the analytic side of the arguments, but we shall also comment on the recent advancements of the causality parts of the theorems \cite{Min:19}. The corresponding  extensions of the singularity theorems in a purely causal setting, as well as in a synthetic setting \cite{GKS:19} and those using methods from optimal transport \cite{CM:20} will be briefly described in the final section \ref{sec:5}. 

To begin with, we discuss the motivation behind this endeavor.

\subsection{Why low regularity}\label{sec:4.1}

Taking up the discussion from Section \ref{sec:2.3} we now take a closer look at the conclusions of the singularity theorems, which---despite all their power and their glory---are generally considered to be a weak spot, cf.\ e.g.\  \cite[Sec.\ 5.1.5]{SG:15}. Indeed, they assert merely the existence of incomplete causal geodesics and, 
%(cf. the discussion in Section \ref{sec:2.3}) 
in general, there is no way to link such a singularity to curvature blow-up in a suitable sense. Also there is the issue of extensions of spacetime: As we have seen, it is essential for the physical interpretation of the theorems that they are applied to (maximally) extended spacetimes\footnote{Some authors include such an assumption into their definition of singular spacetime, see e.g.\ \cite[p.\ 10]{Cla:93}, and, of course, the extensive discussion in \cite[Sec.\ 8.1]{HE:73}.}. 
\medskip

Taking a step back we also see that there is a regularity assumption which is implicit in the classical theorems. In fact, the theorems assert causal geodesic incompleteness of the spacetime \emph{provided} the metric is smooth, and, since the bulk of Lorentzian geometry remains valid there, if it is $C^2$. In particular, they do not exclude the possibility that the spacetime is complete but of lower regularity. 

If the regularity of the metric was just $C^{1,1}$ the curvature would become discontinuous rather than unbounded hence would hardly be considered `singular' on physical grounds. Indeed, via the field equations, this just corresponds to a finite jump in the matter variables. And there are many interesting systems of that type, such as the Oppenheimer-Snyder model of a collapsing star \cite{OS:39}, to give a classical example, and general matched spacetimes, see e.g.\ \cite{Lic:55,Isr:66,MS:93}.
 
Moreover, if the regularity was even lower, one could be inclined to accept such a scenario as long as there is an analytic way to define the curvature and to make sense of the field equations \eqref{eq:ees}. This is indeed possible (in a stable way) for the Geroch-Traschen (or GT-)class of metrics \cite{GT:87}, i.e., metrics of regularity $H^2_{\mbox{{\tiny loc}}}\cap L^\infty_{\mbox{{\tiny loc}}}$ that are uniformly nondegenerate in a suitable sense \cite{LM:07,SV:09}. There the Riemann tensor is a tensor distribution and if a sequence of metrics converges to a GT-regular one in $H^2_{\mbox{{\tiny loc}}}$ then the respective curvatures converge in distributions (for some more details see Section \ref{sec:dcr} below). While current techniques certainly do not carry that far, it would still be interesting to prove singularity theorems in the regularity classes used in the classical existence results for the Einstein equations which is $H^{5/2+\varepsilon}$ or those used in current formulations of the strong cosmic censorship conjecture \cite{Chr:09}, which demand a locally square integrable connection.

However, a first substantial conceptual problem arises with the very notion of geodesic incompleteness: Below $C^{1,1}$ the initial value problem for the geodesic equation is no longer uniquely solvable and  below $C^1$ not even classically meaningful. Thus one would have to resort to some non-classical solution concept, like e.g.\ Filippov solutions \cite{Fil:88}, which have been used in this wider context e.g.\ in \cite{Ste:14,LLS:22}. 
%\todo{geodesics vs. maximisers here?}

A first step in this direction is to lower the differentiability of the metric to $C^1$ for which the curvature is a distribution of order one (again see Section \ref{sec:dcr} below for details) and the i.v.p.\ for the geodesic equation is at least classically solvable if not uniquely so. In fact, we will discuss results in $C^{1,1}$ and $C^1$-regularity below.\medskip

Of course, the regularity issue connected with the singularity theorems was already noted early on and extensively discussed in \cite[Sec.\ 8.4]{HE:73}. There the authors argue that at least the Hawking theorem should continue to hold for $C^{1,1}$-metrics and express their expectation that this should also be true for $C^{0,1}$ (i.e. locally Lipschitz continuous) metrics and also for all the other classical theorems. In fact, a natural next step seems to be to extend the recent $C^{1}$-results to this class and current research is directed at this goal, see also Section \ref{sec:5}.
\medskip

Let us finally come back to the issue of extensions of spacetime. Certainly the classical theorems assert that the incomplete spacetime cannot be extended to a complete one keeping the assumptions and the $C^2$-regularity of the metric. Likewise the results in $C^{1,1}$ and $C^1$ can be read as obstructions to such extensions keeping the respective regularity of the spacetime. This point of view nicely complements recent work by Jan Sbierski \cite{Sbi:18} who showed that the Schwarzschild solution cannot be extended as a continuous\footnote{Note, however, that extensions in even lower regularity do exist, for an overview see \cite{HS:02}.} spacetime. In a similar vein, it has been established in \cite{GLS:18} that timelike geodesic completeness remains an obstruction to extendability also in the class of $C^0$-spacetimes.%, cf.\ Section \ref{sec:2.3}.

\subsection{Low regularity: issues and strategies}\label{sec:4.2}

To begin our technical account, note that the bulk of Lorentzian geometry remains valid for $C^2$-metrics, since the main tools such as normal and convex neighbourhoods as well as normal coordinates are still available. However, slightly below, things begin to worsen gradually. First, the exponential map retains maximal regularity also for $C^{1,1}$-metrics, being a bi-Lipschitz homeomorphism \cite{Min:15,KSS:14}. While this secures the existence of normal and convex neighbourhoods, normal coordinates are of limited use since there the metric is only continuous. Moreover, for metrics of H\"older regularity $C^{1,\alpha}$ for any $\alpha<1$, convexity breaks down completely, since the exponential map needs not be injective on any zero-neighbourhood of the origin as is demonstrated by a Riemannian example in \cite{HW:51}, which easily gives rise to a static Lorentzian example, see \cite{SS:18}. There it is also shown that, in general, causal geodesics fail to locally maximise the Lorentzian distance.
\medskip

Returning to the context of the singularity theorems, let us briefly collect the issues at hand and discuss the strategies we will employ in the sections to come. For a longer and more technically detailed list of places where the classical proofs rest on the $C^2$-differentiability see \cite[Ch.\ 6.1]{Sen:98}.
\begin{enumerate}[label={(\Alph*)}]
 \item\label{A} The curvature is only locally bounded for $g\in C^{1,1}$ and merely a distribution of order one for $g\in C^1$.
 \item\label{B} Normal neighbourhoods are not useful for $g\in C^{1,1}$ and the exponential map is no longer even defined for $g\in C^1$.
 \item\label{C} The geodesic equation fails to be uniquely solvable for $g\in C^1$.
\end{enumerate}
\medskip

Item \ref{A} and, more precisely, the failure of the curvature to be defined pointwise causes several difficulties. First the energy conditions (E) have to be adapted, where especially  \eqref{eq:nec} and \eqref{eq:gc} turn out to be a delicate matter as we will discuss in detail below. Then, Jacobi fields cannot be reasonably defined which, of course, means that one has to do without using the central concepts of conjugate and focal points. A clear strategy to address this issue---as was already pointed out in \cite[Sec.\ 8.3]{HE:73}---is regularisation of the metric by smooth approximations for which the classical tools are still available. We will generally pursue this path and explain it in some detail in the following Section \ref{sec:dcr}. Then in Section \ref{sec:4.4} and in Section \ref{sec:advfoc} we will see how appropriate distributional versions of the energy conditions (E) lead to the focusing of causal geodesics for the approximating metrics.

Next, item \ref{B} makes it necessary to revisit the whole machinery of causality theory and to extended it to the regularity at hand. We will refrain from going into any technical details here and just summarise the necessary background in Section \ref{sec:4.5}

Now, \ref{C} first of all forces us to make a choice concerning the conclusion of the theorems, namely on the notion of incompleteness. It has turned out to be favourable to use the more stringent alternative\footnote{Alternatively one could have only asked for the existence of one complete geodesic for any set of initial data.} put forward in \cite{G:20}:

\begin{definition}[$C^1$-completeness]
A $C^1$-spacetime is called timelike (respectively null or causal) geodesically complete if all inextendible timelike (respectively null or causal) solutions of the geodesic equation are defined on all of $\Reals$.
\end{definition}

However, despite the loss of uniqueness many properties of geodesics extend from the smooth to the $C^1$-setting. In particular, geodesics do have a fixed causal character \cite[Cor.\ 2.3]{G:20} and so the above definition makes sense. 
%What becomes more subtle is the relation between causal geodesics and maximising curves, see e.g.\ \cite{SS:18}.
\medskip

One main issue will be to adjust the following vital aspect of the regularisation approach: We need to approximate maximising causal geodesics of the low regularity metric by maximising causal geodesics of the approximating metrics, which becomes a delicate matter in the absence of an exponential map (B) and unique solvability of the geodesic equation (C). This issue was addressed for globally hyperbolic $C^1$-spacetimes in \cite[Sec.\ 2]{G:20} by establishing that between any pair of points $p<q$ there is (at least) one maximising causal geodesics that can be suitably approximated by a sequence of maximising causal geodesics of approximating metrics. This and corresponding results for points in the chronological future of a Cauchy surface and in the horismos of a closed spacelike codimension $2$ surface, allow for corresponding proofs of the Hawking and the Penrose theorem, which we will sketch in Section \ref{sec:4.6}. However, in the context of the Hawking-Penrose theorem more general results are needed, which will ultimately force us to introduce a new condition, namely a \emph{non-branching} assumption for maximising causal geodesics, to be detailed in Section \ref{sec:4.8}. This will finally allow us to discuss the recent $C^1$-version of the Hawking-Penrose theorem \cite{KOSS:22} in Sections \ref{sec:4.9} and \ref{sec:4.10}.

\subsection{Distributional curvature \& regularisation}\label{sec:dcr}

Here we briefly discuss the general distributional framework in which to understand the curvature of metrics\footnote{All that is said here also applies to metrics of arbitrary signature.} of regularity below $C^2$, with a special emphasis on the $C^1$-case. For more information consult \cite{Mar:68, GT:87,LM:07,GKOS:01,Ste:08}.

The space of (scalar) \emph{distributions of order $k$} on $M$ is the topological dual of the space of compactly supported $C^k$-one-densities $\Gamma^k_c(M,\mathrm{Vol}(M))$ (here $\mathrm{Vol}(M)$ denotes the volume bundle), i.e.,
\[
\Dpk(M) := \Gamma^k_c\big(M,\mathrm{Vol}(M)\big)'.
\]
Similarly, the space of \emph{distributional $(r,s)$-tensor fields of order $k$} is defined as
\begin{equation}
\Dpk\mathcal{T}^r_s(M) \equiv \Dpk(M,T^r_s M) := \Gamma^k_c\big(M,\mathcal{T}^s_r(M) \otimes \mathrm{Vol}(M)\big)'.
\end{equation}
Here $\mathcal{T}^r_s(M)$ denotes the space of smooth tensor fields of rank $(r,s)$, i.e., $\mathcal{T}^r_s(M)=\Gamma^\infty(M,T^r_s)=\Gamma(M,T^r_s)$, since we generally omit $k$ if it is infinite. For the spaces of vector fields, one forms and tensor fields of finite differentiability we will write $\X_{C^k}$,  $\Omega^1_{C^k}$, and $(\mathcal{T}^r_s)_{C^k}$, respectively.  With this notation we have\footnote{The isomorphisms in \eqref{eq:Ck-ext} are algebraic and bornological but not topological, cf.\ \cite{Nig:13}.}
\begin{equation}\label{eq:Ck-ext}
\begin{split}
\Dpk\mathcal{T}^r_s(M) &\cong \Dpk(M) \otimes_{C^k(M)} (\mathcal{T}^r_s)_{C^k}(M)\\ 
&\cong L_{C^k(M)}\Big(\Omega^1_{C^k}(M)^r\times \X_{C^k}(M)^s; \Dpk(M)\Big).
\end{split}
\end{equation}
The first line says that distributional tensor fields are sections of the corresponding tensor bundle with distributional coefficients\footnote{Here $\otimes_{C^k(M)}$ denotes the balanced tensor product over the module $C^k(M)$.}, and the second line reveals them as $C^k$ multilinear maps on one forms and vector fields of regularity $C^k$ that give scalar distributions of order $k$. 
This fact for $k=1$ will turn out to be essential in formulating the genericity condition for $C^1$-metrics below.
\medskip

A \emph{distributional connection} is a map $\nabla: \X(M)\times \X(M) \to \D'\mathcal{T}^1_0(M)$ satisfying 
the usual computational rules: $\nabla_{f X+X'}Y = f\nabla_XY + \nabla_{X'}Y$, $\nabla_X(Y+Y') = \nabla_X Y +\nabla_X Y'$, $\nabla_X(f Y) = X(f)Y + f\nabla_X Y$ ($X,X',Y,Y'\in \X(M)$ and $f\in C^\infty(M)$).
Denoting by ${\mathcal G}$ any of the spaces $C^k$ $(0\leq k)$ or $\lpl$ $(1\leq p)$, we call a distributional connection a \emph{${\mathcal G}$-connection}, if $\nabla_X Y$ is a ${\mathcal G}$-vector field for any $X,Y\in \X(M)$. A particularly important case are $\ltl$-connections since they allow to define the curvature tensor $R: \X(M)^3 \to \D'\mathcal{T}^1_0(M)$ via the usual formula 
\begin{equation}
  R(X,Y)Z:=[\nabla_X,\nabla_Y]Z-\nabla_{[X,Y]}{Z}.
\end{equation}

Moreover, if $E_i$ is a local frame in $\X(U)$ and $E^j\in \Om^1(U)$ is its dual frame, then the Ricci tensor is given by
\begin{equation}\label{eq:Ric_def}
\Ric(X,Y) := (R(X,E_i)Y)(E^i) \in \D'(U).
\end{equation}
The significance of the GT-regular metrics is then rephrased by saying that their Levi-Civita connections actually are $\ltl$-connections. More specifically, metrics $g\in C^1$ have $C^0$-Levi-Civita connections which implies that their Riemann tensor $R\in \Dpo\mathcal{T}^1_3(M)$ as well as their Ricci tensor and scalar curvature are of order $1$. Finally, the standard local formulae hold in $\Dpo$ and in \eqref{eq:Ric_def} we even can use $E_i$ of regularity $C^1$,  which is of great technical importance since it makes it possible to use frames derived via parallel transport w.r.t.\ a $C^1$-metric.
\medskip

As already indicated above our method of choice to deal with the analytic arguments in the proofs of the singularity theorems is regularisation and we will detail our convolution based approach below. Prior, 
we outline our overall strategy to implement an analytic machinery that forces causal geodesics of the rough metric to stop maximising:
%We will base our approach on chartwise convolution of the metric but combine it with the following more sophisticated strategy: 
We will formulate suitable energy conditions for the low regularity metric $g\in C^1$ and derive from it surrogate energy conditions for a sequence of smooth approximating metrics $g_\eps$. These will be weaker than the classical conditions \eqref{eq:sec} and \eqref{eq:nec} in so far as the corresponding expressions\footnote{Here and whenever necessary we will indicate the metric from which a specific quantity is derived using square brackets.} $\Ric[g_\eps](X,X)$ will be shown to be (only) mildly negative. We will then extend the arguments explained in Sections \ref{sec:3.2}, \ref{sec:3.5} in order to still prove the occurrence of focal/conjugate points along causal $g_\eps$-geodesics. (Note, that due to (A) we cannot resort to standard results using averaged energy conditions as put forward e.g.\ in \cite{FG:11}.) This will eventually force the geodesics of $g$ itself to stop maximising. 
To achieve these goals we have to take the following into account:
\begin{enumerate}\renewcommand{\labelenumi}{(R{\arabic{enumi}})}
 \item\label{r1} When deriving the surrogate energy conditions for $g_\eps$ from conditions on $g$ we face the problem that while $\Ric[g_\eps]\to\Ric[g]$ distributionally, we cannot achieve local uniform convergence even for $g\in C^{1,1}$. Therefore we will have to compare $\Ric[g_\eps]$ to a regularisation of $\Ric[g]$ instead.
 \item Since we want to use the classical arguments on smooth approximations $g_\eps$ to derive focusing results, we have to control their causality in terms of the causality of $g$. This can be done thanks to an adapted regularisation procedure put forward by Chrusciel and Grant \cite{CG:12}, that provides us with approximations $\check g_\eps$ and $\hat g_\eps$ with narrower and wider lightcones\footnote{We say that $g_1$ has narrower lightcones than $g_2$ (or $g_2$ has wider lightcones than $g_1$), denoted by $g_1 \prec g_2$, if $g_1(X,X)\leq 0$ implies $g_2(X,X)<0$ for any $X\not=0$.} than $g$, respectively. 
 \item We will have to relate the geodesics of the approximating metrics to the geodesics of $g$. More precisely we will have to show that maximising causal $g$-geodesics are $C^1$-limits of $g_\eps$-causal $g_\eps$-maximising $g_\eps$-geodesics.
\end{enumerate}
\medskip

We proceed by introducing the regularisations to be used in detail. The basic ingredient is  chartwise convolution and we begin by choosing a mollifier, i.e., a smooth, nonnegative 
function $\rho$ on $\Reals^n$, supported in the unit ball, and with unit integral. Then we cover $M$ by a countable and locally finite family of relatively compact chart neighbourhoods $(U_i,\psi_i)$ ($i\in \N$) and let $(\zeta_i)_i$ be a subordinate partition of unity with $\mathrm{supp}(\zeta_i)\subseteq U_i$. % for all $i$. 
Then we choose a family of cut-off functions $\chi_i\in C^\infty_c(U_i)$ with $\chi_i\equiv 1$ on a neighbourhood of $\mathrm{supp}(\zeta_i)$. Finally, for $\eps\in (0,1]$ we set $\rho_{\eps}(x):=\eps^{-n}\rho\left (\frac{x}{\eps}\right)$.
Then, denoting 
%by $f_*$ (respectively $f^*$) 
the push-forward and pull-back of distributions with upper and lower stars, 
%under a diffeomorphism $f$, 
consider for any $\mathcal{T} \in \D'\mathcal{T}^r_s(M)$  the expression
\begin{equation}\label{eq:M-convolution}
\mathcal{T}\star_M \rho_\eps(x):= \sum\limits_i\chi_i(x)\,\psi_i^*\Big(\big(\psi_{i\,*} (\zeta_i\cdot \mathcal{T})\big)*\rho_\eps\Big)(x).
\end{equation}
Here, $\psi_{i\,*} (\zeta_i\cdot \mathcal{T})$ is viewed as a compactly supported distributional tensor field on $\Reals^n$, so componentwise convolution
with $\rho_\eps$ yields a smooth field on $\Reals^n$. The cut-off functions $\chi_i$ secure that $(\eps,x) \mapsto \mathcal{T}\star_M \rho_\eps(x)$
is a smooth map on $(0,1] \times M$. For any compact set and small $\varepsilon$, equation 
%$K\comp M$ there is an $\eps_K$ such that for all $\eps<\eps_K$ and all $x\in K$ 
\eqref{eq:M-convolution} reduces to a finite sum with all $\chi_i\equiv 1$, hence to be omitted from the formula.
%, namely when $\eps_K$ is less than the distance between the support of $\zeta_i\circ\psi_i^{-1}$ and the boundary of $\psi_i(U_i)$ for all $i$ with $U_i\cap K\neq \emptyset$.

Just as is the case for smoothing via convolution in the local setting we obtain optimal convergence, that is  $\mathcal{T}\star_M \rho_\eps$ converges to $\mathcal{T}$ in $C^k$ or $W^{k,p}_{\mbox{{\tiny loc}}}$ $(p<\infty)$ if $\mathcal{T}$ is contained in these spaces\footnote{This, of course, means convergence in the respective norms on compact sets.\label{foot26}}. In particular, for $g\in C^1$ we now set
\begin{equation}
 g_\eps:=g \star_M \rho_\eps\
\end{equation}
to obtain a sequence (actually a net) of smooth Lorentzian metrics that converges\footref{foot26} in $C^1$ to $g$. It is now essential for our purposes to tweak this construction to obtain a regularisation adapted to the causality as indicated above in (R2). We will use the version of \cite[Lem.\ 4.2, Cor.\ 4.3]{G:20}:

\begin{lemma}[Regularisations and convergence]\label{Lemma: approximatingmetrics}
Let $(M,g)$ be a $C^1$-spacetime. Then for any $\eps>0$ there exist smooth Lorentzian metrics $\check{g}_\eps$, $\hat g_\eps$ on $M$ %,time orientable by the same timelike vector field as $g$, and 
satisfying
\begin{equation}\label{eq:geps}
  \check g_\eps \prec g \prec \hat g_\eps\ \mbox{for all $\eps$}\quad\mbox{and}\quad 
  \check g_\eps,\ \hat g_\eps \to  g\ \mbox{in $C^1$}\ (\eps\to 0).
\end{equation}
% 
% \begin{itemize}
% \item[(i)]
% \item[(ii)]  as $\eps\to 0$.
% %\item[(iii)] %$\check g_\eps - g\star_M \rho_\eps \to 0$ in $C^\infty_{\mathrm{loc}}$ and 
% \end{itemize}
%
Moreover we have control on the speed of convergence of $g_\eps$ and compatibility between $\check g_\eps$ and $g_\eps$ as follows: For any compact $K$ there is $c_K>0$ such that for small enough $\eps$
\begin{equation}\label{eq:approximatingmetrics}
 \|g - g_\eps\|_{\infty,K} \leq c_K \eps
 \quad\mbox{and}\quad 
 \|\check g_\eps -g_\eps \|_{\infty,K} \le c_K \eps.
 \end{equation}
An analogous statement holds for $\hat g_\eps$ as well as for the inverse metrics $g^{-1}$, $g_\eps^{-1}$, $(\check g_\varepsilon)^{-1}$, and $(\hat g_\varepsilon)^{-1}$.
\end{lemma}

With these preparations we may now have a look at the distributional energy conditions for metrics $g\in C^1$ and how they imply focusing of causal geodesics for approximating metrics.

\subsection{Distributional energy conditions \& focusing}\label{sec:4.4}
\label{sec:EC}
In this section we introduce the energy conditions akin to \eqref{eq:sec} and \eqref{eq:nec} for $C^1$-metrics and show that they imply suitable surrogate energy conditions on the approximations, cf.\ (R1) above. Here ``suitable'' means that we can still use the surrogate conditions to show focusing results (cf. (R2)) for the geodesics of the approximating metrics---despite the fact that they are weaker than the classical conditions, which are manifestly violated, but only by a controlled margin.

We will be more detailed in case of the \eqref{eq:sec} where we point out the main ideas in the proofs, and will be more sketchy in case of the technically more demanding \eqref{eq:nec}.
\medskip

To begin with, recall that a scalar distribution $u\in\D'(M)$ is nonnegative, $u\geq 0$, if $u(\omega)\geq 0$ for all nonnegative test densities $\omega\in\Gamma_c(M,\mathrm{Vol}(M))$. A nonnegative distribution is always a measure %\cite[Thm.\ 2.1.7]{Hoe90} 
and hence a distribution of order $0$. Moreover, non-negativity is stable with respect to regularisation,\footnote{Recall that we have chosen a mollifier $\rho\geq 0$.} for details see
\cite[Thms.\ 2.1.7 and 2.1.9]{Hoe:03}. 
Finally, for $u,v\in\D'(M)$ we write $u\geq v$ if $u-v\geq 0$. 
\medskip

We say that a $C^1$-spacetime satisfies the \textit{distributional strong energy condition} \eqref{eq:dsec}, 
if  
\begin{equation}\label{eq:dsec}\tag{DSEC}
 \Ric(X,X)\geq 0\ \mbox{in $\mathcal{D}^{'(1)}(M)$ for all timelike $X \in \mathfrak{X}(M)$}.
 %\Ric(X,X)\ \mbox{is nonneagtive in $\mathcal{D}^{'(1)}(M)$ for all timelike $X \in \mathfrak{X}(M)$}.
\end{equation}
The new condition \eqref{eq:dsec} is compatible with the usual classical conditions as well with their obvious reformulation in $L^\infty$ for $g\in C^{1,1}$. The surrogate energy condition for the approximation now is, cf.\ \cite[Lem.\ 4.6]{G:20}:

\begin{lemma}[Surrogate \eqref{eq:sec}]\label{Corollary: timelikeenergyconditionapproxmetrics2}
Let $(M,g)$ be a $C^1$-spacetime satisfying \eqref{eq:dsec} and let $K\subseteq M$ be compact. Then
%$\forall \delta > 0 \ \exists \varepsilon_0 > 0 \ \forall \varepsilon < \varepsilon_0 \ \forall X \in TM\mid_K \text{ with } \check{g}_{\varepsilon}(X,X) = -1$ we have
for all $\delta > 0$ and all smooth $X \in TM\mid_K$ with $\check{g}_{\varepsilon}(X,X) = -1$ we have 
\begin{equation}\label{eq:ssec}\tag{SSEC}
     \Ric[\check{g}_{\varepsilon}](X,X) > - \delta
\end{equation}
for small enough $\eps$. %(i.e. for all $\eps\leq\eps_1$ for some $\eps_1(K,\delta)$).
\end{lemma}

\noindent
\emph{Sektch of proof.} As already remarked above in (R1) the main problem is that the convergence of $\Ric[\check g_\eps]$ to $\Ric[g]$ is not good enough to directly carry the positivity of $\Ric[g](X,X)$ through the argument. Rather we proceed as follows: By standard properties of the convolution  \eqref{eq:dsec} implies $\big(\Ric[g]\star_M \rho_\eps\big)(X,X)>0$ and the result will follow from the compatibility of the distinct regularisations we have used. Indeed we are done, if we can show that $\Ric[g]\star_M \rho_\eps- \Ric[g_\eps]$ and  $\Ric[g_\eps]-\Ric[\check g_\eps]$ both go to zero locally uniformly. 

To establish these statements, we have to estimate the convolution of a product $a\, f$ (basically the components of the inverse metric times a derivative of the components of the metric as occurring in the Christoffel symbols) to a corresponding product of convolutions. After reducing everything to the local situation, this is done by the following Friedrichs-type lemma which takes as an essential input the final estimate in Lemma \ref{Lemma: approximatingmetrics}, i.e., \eqref{eq:approximatingmetrics} for the inverse metrics. Indeed, the components of the inverse of the regularised metric precisely possess the convergence properties assumed for $a_\eps$ in the Lemma below, cf. \cite[Lem.\ 4.9]{G:20}. \hfill\qed

\begin{lemma}[Friedrichs lemma] \label{lem:Fthesecond} Let $f\in C^0(\Reals^n)$ and let $a,a_\eps \in C^1(\Reals^n)$ with $\Vert a_\eps -a\Vert_{\infty}\leq C\eps$ on compact sets. Then $a_\eps (f\star \rho_\eps)- (af)\star \rho_\eps \to 0$ in $C^1$.
\end{lemma}

The next step is to derive an improved focusing result for \emph{smooth} metrics satisfying \eqref{eq:ssec}. Indeed, tweaking somewhat the estimates that led to \eqref{eq:ray-est} one may derive the following result which is a replacement for Proposition \ref{prop:eofp}, cf.\ \cite[Lem.\ 4.10]{G:20}: 

\begin{proposition}[Focal points from negative expansion under \eqref{eq:ssec}]\label{lem:hawkingprep} 
Let $g$ be smooth and let $\gamma:[0,\beta)\to M$ be an inextendible timelike geodesic starting at $p$ normally from a spacelike hypersurface $S$. If $\theta(0)$ is negative (equivalently if the convergence $k(p)$ is positive) and if $\Ric(\dot\gamma,\dot\gamma)\geq -\delta$ for some $\delta<\theta(0)^2/(n-1)$, then there is a focal point $\gamma(b)$ to $S$ for some $b\leq-(n-1)\theta(0)/(\theta(0)^2-\delta (n-1))$, provided $b<\beta$. Consequently $\gamma$ stops maximising the Lorentzian distance to $S$ (the latest) at $b$ if it exists that long.
\end{proposition}

This focusing result will play a mayor role in the proof of the $C^1$-Hawking theorem.
Let us now turn to the corresponding focusing result for the $C^1$-Penrose theorem. First a proper distributional formulation of \eqref{eq:nec} needs some more care due to the fact that vectors that are $\check{g}_{\varepsilon}$-null are only almost $g$-null, cf.\ \cite[Sec.\ 5]{G:20}. 
\medskip

We say that a $C^1$-spacetime satisfies the \textit{distributional null energy condition} \eqref{eq:dnec}, if for any compact set $K$ and any $\delta > 0$ there exists $\varepsilon = \varepsilon(\delta,K) > 0$ such that 
\begin{equation}\label{eq:dnec}\tag{DNEC}
\Ric(X,X) > - \delta \quad\mbox{in $\mathcal{D}^{'(1)}(M)$} 
\end{equation}
for any local smooth vector field $X \in \mathfrak{X}(U)$ ($U \subseteq K$ open) with $\|X\|_h = 1$ and $\mid g(X,X)\mid < \varepsilon$ on $U$. Here $\|\ \|_h$ denotes the norm with respect to some complete Riemannian background metric $h$. Note that this and all future such conditions and estimates will be local and hence in fact be independent of the choice of $h$.
\medskip

Again the new condition \eqref{eq:dnec} is compatible with both \eqref{eq:nec} in the smooth case, and the almost everywhere condition used in the $C^{1,1}$-case. 
The following analogue of Lemma \ref{Corollary: timelikeenergyconditionapproxmetrics2} shows that the above definition of the null energy condition is the correct one in the sense that it produces the following surrogate energy condition on the level of approximations, see \cite[Lem.\ 5.5]{G:20}: 
%Its proof is a technically refined version of the one sketched above for the \eqref{eq:ssec}. 

\begin{lemma}[Surrogate \eqref{eq:nec}]\label{Lemma: nullenergyconditionapproxmetrics}
Let $(M,g)$ be a $C^1$-spacetime satisfying \eqref{eq:dnec}. Let $K\subseteq M$ be compact and let $c_1,c_2 > 0$. Then for all $\delta > 0$ there is $\varepsilon_0 = \varepsilon_0(\delta,K,c_1,c_2) > 0$ such that $\forall \varepsilon < \varepsilon_0 \ \forall X \in TM\mid_K \text{ with } 0 < c_1 \leq \Vert X\Vert_h \leq c_2 \text{ and } \check{g}_{\varepsilon}(X,X) = 0$
\begin{equation}\tag{SNEC}\label{eq:snec}
     \Ric[\check{g}_{\varepsilon}](X,X) > - \delta.
\end{equation}
\end{lemma}

Now the following result is both the generalisation of the arguments that lead to smooth focusing in Propositions \ref{prop:ecp} and \ref{prop:tsts} to the $C^1$-setting, and the null version of Proposition \ref{lem:hawkingprep}. It says that negative expansion still leads to focusing even under \eqref{eq:snec}. We formulate it in a slightly different and quantified manner which assumes the geodesic to be maximising and then restricts its length, for details see \cite[Lem.\ 5.6]{G:20}.

\begin{proposition}[Smooth focusing under \eqref{eq:snec}] \label{lem:penroseprep}
Let $g$ be smooth and let $\gamma:[0,\beta]\to M$ be a future directed null geodesic starting at $p$
from a spacelike submanifold $P$ of codimension $2$. If $\gamma$ is maximising the distance to $P$ (and hence starts normal to $P$) we have: If $\theta(0)=-(n-2)\langle H,\dot{\gamma}(0)\rangle\leq \theta_0<0$ and if $\Ric(\dot{\gamma},\dot{\gamma})\geq -\delta $ with $0\leq \delta \leq \frac{3\theta_0}{4\beta}$, then $\beta\leq -\frac{4(n-2)}{3\theta_0}$.
\end{proposition}

We will return to the issue of focusing in Section \ref{sec:advfoc} where we will also introduce an appropriate distributional version of the genericity condition. First we discuss the $C^1$-versions of the first two of the classical theorems in Section \ref{sec:4.6}. But to do so, we have to briefly turn to causality theory in low regularity in the next section.

\subsection{A brief word on causality theory}\label{sec:4.5}

The regularisation of Lorentzian metrics with controlled causality (cf.\ Lemma \ref{Lemma: approximatingmetrics}) put forward in the seminal paper \cite{CG:12} was in turn used to study causality theory of continuous metrics. In fact, this paper together with \cite{FS:12} initiated the recent systematic study of causality theory in low regularity, see \cite[Sec.\ 1]{Min:19} for a brief overview. 

In this way the bulk of Lorentzian causality theory has been transferred to $C^{1,1}$-spacetimes.
%, where the exponential map and convex neighbourhoods are still available \cite{Min:15,KSSV:14,KSSV:15}. 
While convexity fails below that regularity (cf.\ Section \ref{sec:4.2}) nevertheless most aspects of causality theory can be maintained even under Lipschitz regularity of the metric. Further below some significant changes occur \cite{CG:12,GKSS:20}, while some robust features continue to hold even in more general settings \cite{Min:19,KS:18,BS:18,GKS:19}, see also Section \ref{sec:5} below. 
\medskip

In particular, for $C^1$-spacetimes we may still build the causality relations on local Lipschitz curves\footnote{We could equivalently have used piecewise smooth or $C^1$-curves, for details see \cite[Rem.\ 1.2]{G:20}, and \cite[Sec.\ 2]{GKSS:20} for a general discussion on the choice of classes of curves in low regularity causality theory.}.
Also the push-up principle is still valid and $I^+(A)$ is open for any set $A\subseteq M$. Moreover, even for continuous metrics one may consistently define global hyperbolicity via causality and compactness of causal diamonds $J(p,q)=J^+(p)\cap J^-(q)$,  \cite[Prop.\ 2.20]{Min:19}, which is still equivalent to the existence of a Cauchy surface, 
%(i.e., a set that is met exactly once by every inextendible causal curve), 
see \cite[Thms.\ 5.7 and 5.9]{Sae:16}. Also it is then clear that if $g$ is globally hyperbolic, so is $\check g_\eps$.
\medskip

From there all further ingredients needed in our approach to the singularity theorems in $C^1$ can be derived in $C^1$-regularity, see \cite[Appendix A]{KOSS:22}. These results, however, also follow by compatibility from the more general approaches of \cite{CG:12,Sae:16} and, in particular \cite{Min:19}, where cone structures on manifolds are studied and many results are derived under minimal regularity assumptions.
We will briefly return to this issue in the final Section \ref{sec:5}.
\pagebreak%artificial

\subsection{The Hawking and the Penrose theorems in $C^1$}\label{sec:4.6}

We are now finally ready to present the extensions of the first two of the three classical singularity theorems we have been dealing with throughout. We start with the Hawking theorem which was given for $C^{1,1}$-metrics in \cite[Thm.\ 1.1]{KSSV:15} and generalised to $C^1$ in \cite[Thm.\ 4.11]{G:20} as follows:

\begin{theorem}[$C^1$-Hawking]\label{thm:C1H} Let $(M,g)$ be a  $C^1$-spacetime such that 
\begin{itemize}
 \item [(E)]  the timelike Ricci curvature is nonnegative in $\D'$, i.e., \eqref{eq:dsec} holds, 
 \item [(C)]  there is a compact spacelike Cauchy surface $S$ in $M$, with
 \item [(I)]  everywhere positive future convergence $k$.
\end{itemize}
Then $M$ is future timelike geodesically incomplete.
\end{theorem}

\noindent
\emph{Sektch of proof.} First we assume by contradiction that $(M,g)$ is timelike geodesically complete. Then, by compactness of $S$ and continuity of the mean curvature 
$k:=(n-1)\min_S \langle H,\nu\rangle$ (with $H=H[g]$ and $\nu=\nu[g]$ the future directed $g$-unit normal) exists and is positive. 

By Lemma \ref{Lemma: approximatingmetrics} $\check g_\eps\to g$ in $C^1$ and so $\nu[\check g_\eps]\to \nu$ and $H[ \check g_\eps]\to H$ uniformly on $S$. Therefore $k_\eps:=(n-1)\check g_\eps(H[\check g_\eps],\nu[\check g_\eps])\geq k_0:=k/2 > 0$ for $\eps$ small enough. Also, $\check g_\eps\prec g$ and so $S$ is also $\check g_\eps$-spacelike and all $\check g_\eps$-geodesics starting $\check g_\eps$-normally from $S$ into the future initially have expansion $\theta_\eps(0)\leq -k_0<0$.

Next we want to apply Proposition \ref{lem:hawkingprep} to these geodesics in a uniform way. In order to do so we need to apply \eqref{eq:ssec} with a uniform constant $\delta$ and hence we have to make sure that (the images of) all the $\check g_\eps$-geodesics from above stay in one compact set (at least for small $\varepsilon$), cf.\ Lemma \ref{Corollary: timelikeenergyconditionapproxmetrics2}. To explain in some detail how this is done we introduce the following notation: For any compact $K\subseteq TS$ and any parameter value $t$ we write 
\begin{equation}
 F_{K,t}:=\bigcup\ \{\mathrm{im}(\dot\gamma\mid_{[0,t]})\}\ \subseteq TM,
\end{equation}
where the union runs over all $g$-geodesics with $\dot \gamma(0)\in K$. Similarly we write $F_{\eps,K,t}$ if the $\gamma$-geodesics are replaced by $\check g_\eps$-geodesics with data in $K$.
Now we consider  
\begin{eqnarray*}
K:=&\cup_{0<\eps\leq 1}\{v\in TM\mid_S,\ \mbox{$\check g_\eps$-normal to $S$ and}\ \check g_\eps(v,v)=-1\}\\
   &\cup_{\hphantom{0<\eps\leq 1}}\{v\in TM\mid_S,\ \mbox{$g$-normal to $S$ and}\ g(v,v)=-1\}.
\end{eqnarray*}
and set $b=4(n-1)/k_0$. Then (cf.\ \cite[Prop.\ 2.9]{G:20}) there is $\eps_0(K,b)$ such that 
\begin{equation}
 F_{\leq \eps_0,K,b}:=\bigcup_{0<\eps\leq \eps_0(K,b)} F_{\eps, K,b} \cup F_{K,b}
\end{equation}
is a relatively compact subset of $TM$. 

Denote by $L$ a compact neighbourhood of $F_{\leq\eps_0,K,b}$. Then setting $\delta_0=k_0^2/(2(n-1))$, Lemma \ref{Corollary: timelikeenergyconditionapproxmetrics2} gives us $\eps_1(L,\delta_0)$ such that \eqref{eq:ssec} holds on $L$ with $\delta_0$ and for all $\eps\leq\eps_1$. Therefore, for all $\eps\leq\min(\eps_0,\eps_1)$ we may apply Proposition \ref{lem:hawkingprep} to conclude that the respective $\check g_\eps$-geodesics stop maximising the $\check g_\eps$-distance to $S$ the latest at $b_0=2(n-1)/k_0<b$. 

Therefore, as in the proof of the classical Hawking theorem \ref{thm:CH}, we have that for $\eps$ small $D^+[\check g_\eps](S)$ is contained in a compact set. Recall that $S$ is a Cauchy surface also for $\check g_\eps$. But the Cauchy horizon $H[\check g_\eps](S)$ is nonempty since $\emptyset\not=F_{\eps,K,b}\setminus F_{\eps,K,b_0}\subseteq I^+[\check g_\eps](S)\setminus D^+[\check g_\eps]S$, a contradiction.\hfill\qed
\medskip

Note that we have neither shown nor needed that also $D^+(S)\subseteq\{p\in M:\ d(S,p)\leq 1/k\}$. This fact, however, can be established (cf.\ the discussion preceding \cite[Thm.\ 4.13]{G:20}) using that also for globally hyperbolic $g\in C^1$ maximising geodesics exist. More precisely, the Avez-Seifert theorem was established even for $g\in C^0$ in \cite{Sae:16}, establishing the existence of maximising causal \emph{curves} between any pair of points $p<q$. In $C^{1,1}$- and in $C^1$-spacetimes these maximisers are also geodesics by \cite[Thm.\ 6]{Min:15} and \cite[Thm.\ 3.3]{SS:21}, respectively, but see \cite{LLS:22} for more general results. Hence we have existence of maximising geodesics but as already indicated at the end of Section \ref{sec:4.2}, we also need to approximate these by a sequence of $\check g_\eps$-maximising $\check g_\eps$-geodesics, cf.\ (R3). In the present globally hyperbolic $C^1$-setting this was established for any pair of points $p<q$ and between a Cauchy surface $S$ and any $p\in I^+(S)$ in \cite[Prop.\ 2.12, Cor.\ 2.14]{G:20} (even before \cite[Thm.\ 3.3]{SS:21} was available).

For the proof of the Penrose theorem we will need the following version for null maximisers from $(n-2)$-surfaces given in \cite[Cor.\ 2.15]{G:20}:

\begin{proposition}\label{prop:preppen}
Let $(M,g)$ be a globally hyperbolic $C^1$-spacetime and let $P \sse M$ be a closed, spacelike $(n-2)$-dimensional submanifold. Then for any $q\in E^+(P)$ there exists at least one null geodesic from $P$ to $q$ maximising the distance to $P$. Further, such a geodesic can be obtained  as a $C^1$-limit of a sequence of $\check{g}_{\eps_n}$-null $\check{g}_{\eps_n}$-geodesics $\gamma_{\eps_n}$ maximising the $\check{g}_{\eps_n}$-distance to $P$. 
\end{proposition}

We now proceed to the Penrose theorem which was generalised to $C^{1,1}$-metrics in \cite[Thm.\ 1.1]{KSV:15}
and further to the $C^1$-setting in \cite[Thm.\ 5.7]{G:20} as follows:

\begin{theorem}[$C^1$-Penrose]\label{thm:c1penrose} 
Let $(M,g)$ be a $C^1$-spacetime such that
 \begin{itemize}
 \item [(E)] the null Ricci curvature is nonegative in $\D'$, i.e., \eqref{eq:dnec} holds, 
 \item [(C)] there is a non-compact Cauchy surface $S$, and
 \item [(I)] there is an achronal future trapped surface $P$. 
\end{itemize}
Then $M$ is future null geodesically incomplete.
\end{theorem}

\noindent
\emph{Sketch of proof.} We assume null completeness and first establish that the future horismos $E^+(P)$ of $P$ is compact. Since $P$ is trapped, $\langle H,\nu\rangle>0$ for all future pointing null normals $\nu$. Defining the compact set $K=\{X\in TP^\perp:\ 0\leq \langle H,X\rangle \leq 2 \}$ we show that  $E^+(P)\sse F_{K,1}$ which is compact. This gives the claim since by global hyperbolicity $E^+(P)$ is also closed. 

So let's assume by contradiction that there is $p\in E^+(P)\setminus F_{K,1}$. Since $p\in E^+(P)$, by Proposition \ref{prop:preppen} it is reached by a normal maximising null geodesic $\gamma :[0,1]\to M$ which  is the uniform $C^1$-limit of a sequence of $\check{g}_{\eps_k}$-null geodesics $\gamma_{\eps_k}:[0,1]\to M$ maximising the $\check{g}_{\eps_k}$-distance to $P$. 
Since $p\not\in F_{K,1}$ we have $\langle H,\dot\gamma(0)\rangle > 2$ and so for $k$ large $-(n-2)\check{g}_{\eps_k}(H_{\eps_k},\dot{\gamma}_{\eps_k}(0))<-2(n-2)=:\theta_0$.
By convergence we may choose a compact set and constants $c_1$, $c_2$ to apply Lemma \ref{Lemma: nullenergyconditionapproxmetrics} to obtain \eqref{eq:snec} for all large enough $k$ with an appropriate choice of $\delta<-3\theta_0/4$. 
%we have 
%\begin{equation}
% \Ric[\check{g}_{\eps_n}](\dot{\gamma}_{\eps_n},\dot{\gamma}_{\eps_n})\geq -3\beta \frac{1}{4}=:-3\beta (1-c)
%\end{equation}
Then Proposition \ref{lem:penroseprep} gives $1=\beta\leq -\frac{4(n-2)}{3\theta_0}=\frac{2}{3}$, a contradiction.
\medskip

So, $\partial I^+(P)$, which by global hyperbolicity equals $E^+(P)$, is compact and we aim at topological argument as in the classical proof. To this end we choose a smooth metric $g'\prec g$. Then $S$ is a non-compact Cauchy surface  also for $(M,g')$. 
Further, $I^+(P)$ is a $g'$-future set by $I^+[g'](I^+(P))\sse I^+(I^+(P))= I^+(P)$.  So its boundary $\partial I^+(P)$ is a compact $g'$-achronal topological hypersurface. Now we proceed exactly as in the proof of \ref{thm:CP} to obtain a homeomorphism between $S$ and $\partial I^+(P)$. The latter set is compact while the former is not, again a contradiction.\hfill\qed
\medskip

In the following we begin to turn towards the low regularity extension of the third of the classical theorems, namely the one of Hawking-Penrose. We start with the causal result, i.e., Lemma \ref{lem:hp}. Here we have, of course, to remove the explicit occurrence of conjugate points in condition (C2). However, as we have seen in our previous discussions, what really matters is that inextendible causal geodesics stop maximising the Lorentzian distance. Moreover, a closer inspection of the classical proofs reveals that in the timelike case this property is only needed on open globally hyperbolic subsets of spacetime (cf.\ also the discussion at the end of Section \ref{sec:4.8}). Using the extensions of causality theory briefly discussed in Section \ref{sec:4.5} one may establish the following result, see \cite[Thm.\ 6.2]{KOSS:22}: 

\begin{lemma}[$C^1$-Hawking-Penrose Lemma]\label{thm:C1HPcausalityversion}
In any $C^1$-spacetime $(M,g)$ the fol\-lo\-wing statements cannot simultaneously hold:
\begin{enumerate}
    \item[\ (C1\,)] \label{lem:C1HP1} $M$ is chronological,\footnote{While \cite[Thm.\ 6.2]{KOSS:22} uses causality rather than chronology, a closer look at the proof reveals that this weaker condition suffices. However, in the analytical result we will anyways need causality, cf.\ Theorem \ref{Theorem: HPC1}.} i.e., it contains no closed timelike curve.
    \item[(C2')] \label{lem:C1HP2a} No inextendible timelike geodesic in an open globally hyperbolic subset is maximising.
    \item[(C2'')] \label{lem:C1HP2b} No inextendible null geodesic is maximising.
    \item[(C3\,)] \label{lem:C1HP3} There is trapped set $A$.
\end{enumerate}
\end{lemma}

While in establishing the $C^1$-Hawking-Penrose theorem we will follow the general layout of the classical arguments used in Section \ref{sec:3.5}, the steps from Lemma~\ref{thm:C1HPcausalityversion} to the analytic result is now considerably more involved. Considering the sophisticated analysis of the influence of curvature on causal geodesics needed in the classical proof (cf.\ Proposition \ref{prop:ecp2}) this is not surprising at all: The technical issues arising form the lack of a suitable concept of conjugate points become more pronounced here. In fact, it will take us three more technical sections to prepare the statement and a sketch of proof which we will finally provide in Section \ref{sec:4.10}\footnote{We advise readers mainly interested in the bigger picture to directly jump there.}.
We start our account by discussing the distributional genericity condition in the next Section \ref{sec:advfoc}. We will show that it allows to still derive a focusing result for the causal geodesics of approximating smooth metrics. Then, following our general plan, we want to establish that this leads to focusing of the causal geodesics of the $C^1$-metric $g$. In particular, we have to establish that, under appropriate assumptions, conditions (C2') and (C2'') of Lemma~\ref{thm:C1HPcausalityversion} hold. While this can be achieved using the methods discussed so far in the case of a $C^{1,1}$-metric (cf.\ \cite[Thms.\ 5.1, 5.3]{GGKS:18}), it needs a more careful analysis of how in the $C^1$-case maximising causal geodesics can be approximated by respective geodesics of the approximating metrics (R3). We will discuss this issue in Section \ref{sec:4.8} where we introduce a new condition that prohibits the branching of maximal causal geodesics for $g$. Finally, in Section \ref{sec:4.9} we will deal with appropriately generalised initial conditions and discuss how they each---once again using focusing---lead to the formation of a trapped set.

% We show that appropriate versions of the curvature conditions~\eqref{smoothstrongenergy} and~\eqref{genericityOriginal} lead to causal geodesics becoming non-maximising between their endpoints. We prove this result by studying appropriate smooth approximations $g_{\eps}$ to the $C^{1, 1}$-metric $g$, where the $g_{\eps}$ satisfy appropriate weakened versions of~\eqref{smoothstrongenergy} and~\eqref{genericityOriginal}. By a refined analysis of the matrix Riccati equation along geodesics with respect to the
% $g_{\eps}$-metrics, we are able to show that $g_{\eps}$-causal geodesics develop conjugate points,%
% \footnote{Note that the metrics $g_{\eps}$ are smooth, so the classical notion of a conjugate point is well-defined.} %
% and, hence, are non-maximising. From this, we argue that $g$-causal geodesics also become non-maximising. At this point, our main results, Theorem~\ref{HPinC1,1} and~Theorem~\ref{ArbitrCodiminC1,1} follow from Theorem~\ref{C11HPCausalitybit}.
% 
% The techniques that we develop in going from Theorem~\ref{C11HPCausalitybit} to Theorem~\ref{HPinC1,1} and~Theorem~\ref{ArbitrCodiminC1,1} are the main technical developments in this paper. In particular, the estimates that we develop in Sections~\ref{reg_sec} and~\ref{sec:conjugate_points} are new,%
% \footnote{To the best of our knowledge.}
% and may well be of independent interest.%
% \footnote{In particular, these are~\emph{not\/} estimates that follow from the standard Rauch comparison theorem for Jacobi fields.}

\subsection{Distributional genericity \& advanced focusing}\label{sec:advfoc}

We now wish to introduce an appropriate distributional version of the genericity condition which will allow us to prove the advanced focusing results needed to establish the $C^1$-Hawking-Penrose theorem. The general strategy is the same as employed in Section \ref{sec:4.4}: From the distributional genericity condition we will establish estimates on the tidal force operator for the approximations, which are still strong enough to produce pairs of conjugate points along causal geodesics of the approximating metrics that are ``long enough''. 

As in the case of the \eqref{eq:dnec} we will have to use extensions of vector fields to small neighbourhoods. However, since in the course of our later arguments we are bound to use $g$-frames which possess mere $C^1$-regularity we need to formulate the condition for $C^1$-fields. It is here that we crucially rely on the fact that the Ricci tensor is a first order distribution and hence allows to insert $C^1$-vector fields, cf.\ \eqref{eq:Ck-ext}.
\medskip

We say that along a causal geodesic $\gamma$ the \textit{distributional genericity condition} \eqref{eq:dgc} holds at $\gamma(t_0)$,  if it possesses a neighbourhood $U$ with $C^1$-vector fields
\begin{equation}\nonumber
 \mbox{$X$ restricting to $\dot{\gamma}$, and $V$ restricting to a vector field normal to $\dot\gamma$}
\end{equation}
at all $\gamma(t) \in U$ and there exist $c > 0$ and $\delta > 0$ such that for all $C^1$-vector fields $\tilde{X}$, $\tilde{V}$ on $U$ with $\|X - \tilde{X}\|_h < \delta$ and $\|V - \tilde{V}\|_h < \delta$ we have
\begin{equation}\label{eq:dgc}\tag{DGC}
    g(R(\tilde{X},\tilde{V})\tilde{V},\tilde{X}) > c \quad \text{in } \mathcal{D}^{'(1)}(U).
\end{equation}

This condition again is consistent with its smooth and $C^{1,1}$-counterparts. Also it implies useful estimates on the tidal force operators of the approximations. The precise results are \cite[Lem.\ 2.16, 2.17]{KOSS:22}, which we summarise next. At the heart of the technical proofs is again positivity of the direct convolution of the positive term in \eqref{eq:dgc} and the compatibility of the various regularisations, established by the Friedrichs lemma \ref{lem:Fthesecond}.

\pagebreak%artificial

\begin{lemma}[Bounds on the tidal force operator from \eqref{eq:dgc}]\label{Lemma: genericityfriedrichs}
Let $\gamma$ be a causal geodesic in a $C^1$-spacetime and assume \eqref{eq:dgc} at $\gamma(0)$. 
%Write $g_{\varepsilon}$ for either $\check{g}_{\varepsilon}$ or $\hat{g}_{\varepsilon}$. Further 
Let $\check \gamma_\varepsilon$ be $\check g_{\varepsilon}$-geodesics, whose $\check g_{\varepsilon}$-causal character is the same as the $g$-causal character of $\gamma$ and which converge in $C^1$ to $\gamma$. 
Then there are vector fields $E_i^\eps$ on some neighbourhood $U$ of $\gamma(0)$ such that
\begin{enumerate}
 \item[(i)] $E_i^\eps \circ \gamma_\eps$ is a $g_\eps$-orthonormal frame along $\gamma_\eps$, and
 \item[(ii)] $E_i^\eps \to E_i$ in $C^1(U)$, where $E_i\circ \gamma$ is an orthonormal frame along $\gamma$.
\end{enumerate}
Finally, there are $\tilde c>0$, $r>0$, and $C=C(\eps)>0$ such that along $\gamma_\eps$ the 
%$\eps$-
tidal force operator for small enough $\eps$ satisfies 
%$[R_\eps](t):= [R_\eps(.,\dot{\gamma}_\eps(t))\dot{\gamma}_\eps(t)]:[\dot{\gamma}_\eps(t)]^\perp \to [\dot{\gamma}_\eps(t)]^\perp$ fulfills  
\begin{equation}\label{eq:coest}
 [R[\check g_\varepsilon]](t) > \mathrm{diag}(\tilde c,-C,\ldots,-C) \text{ on } [-r,r].
\end{equation}
\end{lemma}

The announced focusing result asserting the existence of a pair of conjugate points along causal geodesics of smooth metrics given an estimate of the form \eqref{eq:coest} is \cite[Lem.\ 4.1]{KOSS:22}\footnote{This result, which is based on \cite[Prop.\ 4.2]{GGKS:18} uses Riccati comparison techniques that might be interesting in their own right as they are independent of the standard Rauch comparison theorem for Jacobi fields.}:

\begin{proposition}[Advanced smooth focusing]\label{Lemma: conjpointssmoothmetric}
    Let $g$ be smooth. Given some $c>0$ and $0<r< \frac{\pi}{4 \sqrt{c}}$,
    there exist
    $\delta(c,r)>0 $ and $T(c,r)>0$ such that any causal geodesic $\gamma$ defined on $[-T,T]$ for which
    the following hold
    \begin{itemize}
        \item[(i)] $\text{Ric}(\dot{\gamma},\dot{\gamma})\geq - \delta $ on $[-T,T]$, and 
        \item[(ii)] there exists a smooth parallel orthonormal frame for $[\dot{\gamma}]^\perp$ and some
        $C>0$ such that w.r.t.\ this frame the tidal force operator satisfies 
        $$[R](t) > \mathrm{diag}(c,-C,\ldots,-C)\quad\mbox{on}\quad [-r,r],$$
    \end{itemize}
    possesses a pair of conjugate points on $[-T,T]$.
\end{proposition}

\noindent
\emph{Sketch of proof.} We proceed indirectly, assuming that for any $\delta>0$ and $T>0$ there is some $\gamma$ satisfying (i) and (ii) without conjugate points in $[-T,T]$. Denote by $[A]$ the unique Jacobi tensor class along $\gamma$ with $[A](-T)=0$ and $[A](0)=\mathrm{id}$. Writing all linear endomorphisms of $[\dot \gamma]^\perp$ in a basis as in (ii) we set $[\tilde R](t):=\diag(c,-C,\dots,-C)$ so that it is bounded above by $[R](t)$ on $[-r,r]$.

Recall from Section \ref{sec:3.2} that  $[B]:=[\dot A]\cdot [A]^{-1}$ satisfies the matrix Riccati equation 
\begin{equation}\tag{\ref{eq:riccati}} 
[\dot B] + [B]^2 + [R] = 0, 
\end{equation}
and we denote by $[\tilde B]$ the solution to~\eqref{eq:riccati}, with $[\tilde R]$ instead of $[R]$.
Now, the trick is to chose an appropriate initial condition for $[\tilde B]$ at some $t_1\in [-r,0]$ such that $[\tilde B](t_1)\geq[B](t_1)$ since then the comparison theorem of \cite{EH:90} implies $[B]\leq[\tilde B]$ on $[t_1,r]$.

One may actually find such an initial condition in the form $[\tilde B](t_1)=\tilde \beta(t_1)\cdot \mathrm{id}$, where $\tilde \beta(t_1)$ is bounded below by the largest eigenvalue of $[B](t_1)$. This is done via an analysis of the Raychaudhuri equation \eqref{eq:ray} for $\theta=\tr([B])$ which, in the absence of conjugate points on $[-T,T]$, allows to bound $\mid\theta\mid$ on $[-r,r]$ (see \cite[Lem.\ 4.1, 4.2]{GGKS:18} for details). 
\medskip

Finally, since both $[\tilde R]$ and $[\tilde B](t_1)$ are diagonal, the Riccati
equation for $[\tilde B]$ decouples and can be explicitly solved by
\[
[\tilde B](t) = \frac{1}{d}\diag(H_{c,f}(t),H_{-C,f}(t),\dots,H_{-C,f}(t)),
\]
where
\[
H_{c,f}(t) = d \sqrt{c}\cot(\sqrt{c}(t-t_1) + \mathrm{arccot}(f/\sqrt{c})),
\]
and $H_{-C,f}$ is of the same form with the hyperbolic tangent replacing the cotangent.
%$\tanh$ and $\mathrm{artanh}$ replacing $\cot$ and $\mathrm{arccot}$, respectively.
% \[
% H_{-C,f}(t)=d \sqrt{C}\tanh\big(\sqrt{C}(t-t_1)+\mathrm{artanh}(f/\sqrt{C})\big).
% \]
Now again using the Raychaudhuri equation for $\theta$ and explicitly analysing the function $H_{c,f}$ it turns out that under the sole condition $4r\sqrt{c}<\pi$, the constants $\delta$ and $T$ (depending only on $c$) can be chosen small, respectively large enough to arrive at a contradiction to $[B]\leq[\tilde B]$. 
\hfill\qed
\medskip

Now, as discussed at the end of Section \ref{sec:4.6}, we want to establish via the above Proposition that also geodesics for the $C^1$-metric stop maximising if they become too long. We do so in the next section introducing some new ideas.

\subsection{Geodesic branching}\label{sec:4.8}

To motivate the introduction of the new non-branching condition we wish to sketch the essential argument that allows us to 
\begin{eqnarray}\nonumber
 &&\mbox{pass from Prop.\ \ref{Lemma: conjpointssmoothmetric} to the fact that no causal geodesic of $g\in C^1$}
 \\ \label{nb-motivation}
 &&\mbox{is globally maximising under \eqref{eq:dgc} and \eqref{eq:dsec} resp.\ \eqref{eq:dnec}.}
 %\\\nonumber &&\mbox{}
\end{eqnarray}
In fact, we will only do so for timelike geodesics and $g\in C^1$ globally hyperbolic. 
%while this argument goes though in the $C^{1,1}$-case (cf. \cite[Thm.\ 5.1]{GGKS:18}) it fails in $C^1$.
Assume indirectly that we are given a complete timelike geodesic $\gamma$ that is maximising between any of its points. The idea is to first construct a sequence of maximising timelike $\check g_\eps$-geodesic $\gamma_\eps$ from $\gamma(-T)$ to $\gamma(T)$ for some suitably large $T$ which converges to $\gamma$ in $C^1$. Then assuming \eqref{eq:dgc} at $\gamma(0)$ and \eqref{eq:dsec} we want to employ Proposition \ref{Lemma: conjpointssmoothmetric} to show that actually $\gamma_\eps$ cannot be maximising. This allows to reach a contradiction, since the $C^1$-limit $\gamma$ of $\gamma_\eps$ was assumed to be a maximiser.

More precisely the $C^1$-convergence of the approximating geodesics will allow us to apply Lemma \ref{Lemma: genericityfriedrichs} 
to conclude from \eqref{eq:dgc} that \eqref{eq:coest} holds, which actually is assumption (ii) in Proposition \ref{Lemma: conjpointssmoothmetric} for $\check g_\eps$. Furthermore $C^1$-convergence of $\gamma_\eps$ will also allow us to choose an appropriate compact set in $TM$ to turn \eqref{eq:dsec} into assumption (i) of  Proposition \ref{Lemma: conjpointssmoothmetric} for $\check g_\eps$ via Lemma \ref{Corollary: timelikeenergyconditionapproxmetrics2}.
Then for an appropriate choice of $\delta$ (small) and $T$ (large) Proposition \ref{Lemma: conjpointssmoothmetric} asserts that $\gamma_\eps$ has conjugate points and hence cannot be maximising, which gives the desired contradiction.
\medskip

So everything boils down to construct a sequence of maximising timelike $\check g_\eps$-geodesic $\gamma_\eps$ from $\gamma(-T)$ to $\gamma(T)$ that converges to $\gamma$ in $C^1$. In case $g\in C^{1,1}$, this can actually be achieved in the following way: By global hyperbolicity of $\check g_\eps$ there is a maximising timelike $\check g_\eps$-geodesic $\gamma_\eps$ from $\gamma(-T)$ to $\gamma(T)$. Suitably reparametrising $\gamma$ and $\gamma_\eps$ we may achieve that $\gamma(-T)=\gamma_\eps(-T)$ and that $\dot\gamma(-T)$ and $v_\eps:=\dot\gamma_\eps(-T)$ have the same $h$-norm bound. Consequently $v_\eps$ has a subsequence $v_{\eps_k}$ converging to some $w$ with $\| w\|_h=\|\dot\gamma(-T)\|_h$. Therefore $\gamma_{\eps_k}$ converges in $C^1$ to a $g$-geodesic $\gamma_{w}$ which is the unique solution to the geodesic equation with initial data $\gamma_w(-T)=\gamma(-T)$ and $\dot \gamma_w(-T)=w$ which, as a limit of maximisers, is maximising and can be shown to reach $\gamma(T)$. Now if $\gamma_w\not=\gamma$ then both are maximising from $\gamma(-T)$ to $\gamma(T)$ and hence $\gamma$ is not maximising beyond $T$, which contradicts our assumption. Hence we may assume that $\gamma_w=\gamma$ and we have constructed the desired sequence of approximating geodesics, which allows us to follow the arguments laid out above.

However, if $g$ is merely a $C^1$-metric uniqueness of solutions of the geodesic equation fails and the 
above construction does not produce a sequence of approximating geodesics. So we have to look for an alternative. 
\medskip

First, note that in the present situation, i.e., given an ODE with merely continuous r.h.s.\ and possibly different solutions to the same initial data, the following issue arises in \emph{any} approximation approach: Regularising the coefficients of the equation results in a smooth situation with unique solvability of the initial value problem. Now to approximate a given (non-unique) solution of the original i.v.p.\ by (unique) solutions of the regularised problem seems unfeasible unless a specific regularisation is constructed to force exactly the desired convergence. Such an approach is of course completely unsuited to the situation at hand, where we want to approximate a given but \emph{arbitrary} maximising geodesic.

Instead we will employ a \emph{non-branching condition} for maximising geodesics, which is
well motivated by similar conditions used in (Riemanian) \emph{metric geometry}. There, in absence of a differentiable structure, geodesics are defined as (local) minimisers of the length functional and local uniqueness of geodesics is expressed in the form of a {non-branching} condition. Here, a branch point is defined as an element of a minimiser at which the curve splits into two minimisers that on some positive parameter interval do not have another point in common, cf., e.g., \cite{Shi:93,Vil:09}. Similarly, in the synthetic Lorentzian setting \cite{KS:18}, the role of causal geodesics is taken on by maximising causal curves, and non-branching is formulated analogously. In both cases, lower synthetic sectional curvature bounds (formulated via triangle comparison in constant curvature model spaces) imply non-branching \cite{Shi:93,KS:18}. 

While in the present $C^1$-setting, the coincidence between causal local maximisers and geodesics, that is familiar from smooth Lorentzian geometry ceases to hold (\cite[Ex.\ 3.2]{KOSS:22}, \cite{SS:18}), it still seems reasonable to assign a privileged role to causal geodesics that \emph{are} locally maximising and to preclude them from branching. We now explicitly introduce our (rather weak) non-branching conditions, where we do not require the second branch to be maximising as well:

\begin{definition}\label{def:branching} (Non-branching conditions) A geodesic $\gamma:[a,b]\to M$ branches at $t_0\in (a,b)$ if there exits $\eps>0$ and some geodesic $\sigma$ with 
\begin{equation}
 \gamma\mid_{[t_0-\eps, t_0]}\,\subseteq \sigma\quad \mbox{but}\quad 
 \gamma\mid_{(t_0,t_0+\eps)}\cap\ \sigma= \emptyset.
\end{equation}
A $C^1$-spacetime is called maximally causally (resp.\ timelike, resp.\ null) non-branching (MCNB, MTNB, MNNB), if no maximal causal (resp.\ timelike, resp.\ null) geodesic branches in the above sense.
\end{definition}

A $C^1$-spacetime in which maximal causal branching occurs can be found in \cite[Ex.\ 3.2]{KOSS:22}. 
\medskip

We may now finally proceed to establish the existence of appropriate approximating sequences of  maximising geodesics, which is the missing link in our argument for \eqref{nb-motivation}. We only state the result in the timelike case, for the corresponding null-version in which we have to avoid the use of global hyperbolicity see \cite[Prop.\ 34(ii)]{KOSS:22}. The proof is based on the fundamental fact that non-branching prevents distinct maximising geodesics from intersecting tangentially in the interior of their domain, cf.\ \cite[Lem.\ 3.3]{KOSS:22}, and an ODE-argument for equations with continuous right hand side \cite[Ch.\ 2, Thm.\ 3.2]{Har:02}, see also \cite[Cor.\ 2.6]{KOSS:22}.

\begin{proposition}[Approximating geodesics in non-branching spacetimes]\label{prop:approxmcnb}
    Let $(M,g)$ be a globally hyperbolic MTNB $C^1$-spacetime.
 %       Suppose that $M$ is globally hyperbolic and let $\varepsilon_k\searrow 0$ ($k\to\infty$). Set $g_k := \hat g_{\eps_k}$ or $g_k := \check g_{\eps_k}$ for all $k$. 
        If $\gamma:[0,a]\to M$ is a maximising, timelike $g$-geodesic
        then for any small $\eta>0$ there exists a subsequence $\check g_{\eps_k}$ of $\check g_\eps$ and maximising, timelike $\check g_{\eps_k}$-geodesics $\gamma_k$ converging in $C^1$ to $\gamma\mid_{[0,a-\eta]}$.
%         \item Let $g_k = \check g_{\eps_k}$ for all $k$. If $M$ is causal
%         and $\gamma:[0,a]\to M$ is a $g$-maximizing null geodesic, then for
%         any small $\delta>0$, there exists a subsequence $g_{k_l}$ of
%         $g_k$, $t_l\downarrow 0$ and $g_{k_l}$-null geodesics
%         $\gamma_l:[t_l, a-\delta]\to M$ contained in $\partial
%         I_l^+(\gamma(0))$ (hence in particular
%         $g_{k_l}$-maximising), which converge in $C^1_{\mathrm{loc}}$ to
%         $\gamma\mid_{[0,a-\delta]}$.
%         
%          
%         
%         If, furthermore, $M$ is strongly causal and $S$ is an acausal set such that
%         $\gamma \subseteq E^+(S)$, then even $\gamma_l:[t_l,a-\delta]\to E^+_l(S)$ and $\gamma_l(t_l) \in S$ with $\gamma_l(t_l)\to \gamma(0)$.
%     \end{enumerate}
\end{proposition}

Now applying essentially the argument from the beginning of this section (and its corresponding null version) we may establish the nonexistence of inextendible maximising geodesics---commonly called \emph{lines}---in non-branching $C^1$-spacetimes under genericity and the energy conditions. More precisley we have:

\begin{theorem}(No lines)\label{Theorem:nolines}\label{thm:no-lines} 
Let $(M,g)$ be a $C^1$-spacetime satisfying \eqref{eq:dgc}.
\begin{enumerate}
 \item[(i)] If $M$ is globally hyperbolic and MTNB and if \eqref{eq:dsec} holds, then there is no complete timelike line.
 \item[(ii)] If $M$ is causal and MNNB and if \eqref{eq:dnec} holds, then there is no complete null line.
\end{enumerate}
\end{theorem}

We remark on the following subtleties concerning the distinction between the timelike and the null case in the above result: We may assume the (very strong) condition of global hyperbolicity in the timelike case, since Theorem \ref{thm:no-lines}(i) enters the proof of the $C^1$-Hawking-Penrose theorem only via condition (C2') of the causal result \ref{thm:C1HPcausalityversion}\footnote{This is similar to the classical proof where its smooth counterpart Proposition \ref{prop:ecp2}---which, however, holds globally---enters only towards the end, when one already works in some Cauchy development.}. 
However, assuming global hyperbolicity in the null case would render such a statement (although easily proved to hold) mostly useless because inextendible yet maximising null curves need to be excluded everywhere in the spacetime and not just in some globally hyperbolic subset. In particular, it is needed when ``upgrading'' the causality property of the spacetime to strong causality via the exclusion of null lines, cf.\ the final argument establishing Proposition \ref{prop:3.13} for the classical argument, which will be used in the same way below in the $C^1$-proof.

But, fortunately in the null case there is a sharper distinction between maximising and non-maximising geodesics because a null geodesic stops maximising if and only if it leaves the boundary of a lightcone, and we have already exploited the structure of such boundaries, cf.\ the proof of Theorem \ref{thm:CP}.  However, the methods needed in the proof of the above Theorem \ref{Theorem:nolines}(ii) fail for closed null curves since they are badly behaved with respect to approximation, cf.\ proof of Thm.\ 5.3 in \cite{GGKS:18}. So these have to be excluded in the statement by assuming the spacetime to be \emph{causal instead of merely chronological}, as was sufficient in the classical results.

\subsection{Initial conditions}\label{sec:4.9}

In this last of the preparatory sections we generalise the four distinct initial conditions of the Hawking-Penrose Theorem to the present $C^1$-setting. More precisely, we provide suitable extensions of the four initial conditions (I1)--(I4) of the classical theorem \ref{thm:CHP} and discuss how they each lead to the formation of a trapped set, i.e., condition (C3) of the $C^1$-Hawking-Penrose Lemma \ref{thm:C1HPcausalityversion}.
\medskip

The easiest case to deal with is the first, i.e., (I1), since it needs no special attention: Given the extension of causality theory discussed in Section \ref{sec:4.5} it follows just as in the smooth case (cf.\ the end of Section \ref{sec:3.5})  that a compact achronal set $P$ without edge is a topological hypersurface and $E^+(P)=P$, which gives (C3).
\medskip

Next we deal with the trapped submanifold-cases, i.e., (I2) and (I3). In the present regularity class it is natural to extend the corresponding initial conditions to trapped $C^0$-submanifolds defined in the \emph{support sense}.

\begin{definition}(Closed trapped $C^0$-submanifolds)
A compact without boundary, spacelike $C^0$-submanifold $P\subseteq M$ of codimension $1 < m<n$ 
is called a \textit{future trapped submanifold} if 
\begin{enumerate}
 \item[(1)] any point $p\in P$ possesses a neighbourhood $U$ with $U\cap P$ achronal in $U$, and
 \item[(2)] the mean curvature vector field is past pointing timelike in the sense of support submanifolds on all of $P$.
\end{enumerate}
\end{definition}
Condition (2) means that for any $q \in P$ there exists a future $C^2$-support submanifold $\tilde{P}$ for $P$ at $q$ whose mean curvature vector at $q$ is past-pointing timelike. Here a \textit{future support submanifold} $\tilde{P}$ for $P$ at $q \in P$ is a submanifold of the same dimension containing $q$ and such that there is a neighbourhood $U$ of $q$ in $M$ such that $\tilde{P} \cap U \subseteq J^+_U(P)$, the causal future of $P$ within $U$.

Now, to show that, given a closed trapped $C^0$-submanifold $P$, a trapped set forms, i.e. that (C3) holds, we have to again establish that lightrays from $P$ stop maximising, now under \eqref{eq:dnec} (in case $m=2$) and a suitable distributional generalisation of condition \eqref{eq:gs} (in case of general $m$). This condition is very much in the spirit of \eqref{eq:dgc}, in the sense that it also asks for a $C^1$-stability of the condition on the curvature: We consider a future directed normal null geodesic $\gamma$ starting at a point $p\in P$. Choose a frame $e_1(p),\dots,e_{n-m}(p)$ of $T_pP$ and denote by $E_1,\dots E_{n-m}$ ist ($C^1$-)parallel transport along $\gamma$. Now we assume that for any $b$ in the domain of $\gamma$ there is a neighbourhood $U$ of $\gamma\mid_{[0,b]}$ and $C^1$-extensions $\overline{E}_i$ of $E_i$ and $\overline{N}$ of $\dot{\gamma}$ to $U$ such that for each $\delta >0$ there exists $\eta > 0$ such that for all collections of $C^1$-vector fields $\{\tilde{E}_1,\dots,\tilde{E}_{n-m},\tilde{N}\}$ on $U$ with $\|\tilde{E}_i -
\overline{E}_i\|_h < \eta$ for all $i$ and $\|\tilde{N}-\overline{N}\|_h < \eta$, we have
\begin{align}\label{eq:initialcurvature}
    \sum_{i=1}^{n-m} g(R(\tilde{E}_i,\tilde{N})\tilde{N},\tilde{E}_i) \geq -\delta \quad \text{in } \mathcal{D}^{'(1)}(U).
\end{align}
This condition leads to a surrogate version for smooth approximating metrics which, just as \eqref{eq:snec}, secures focusing of approximating geodesics, cf.\ \cite[Props.\ 5.5, 5.7]{KOSS:22}. From there we apply the machinery of Section \ref{sec:4.8} to ``lift'' the result to MNNB $C^1$-spacetimes. Finally, we obtain \cite[Prop.\ 5.10, 5.11]{KOSS:22}:

\begin{proposition}[Trapped set from trapped $C^0$-submanifold]\label{Proposition:submfdnullcompletenessortrappedset}
Let $(M,g)$ be a strongly causal, MNNB $C^1$-spacetime and let $P\subseteq M$ be a trapped $C^0$-submanifold of codimension $1<m<n$. In case
\begin{enumerate}
 \item[(i)] $m = 2$ suppose \eqref{eq:dnec}, and in case 
 \item[ii)] $2<m<n$ suppose \eqref{eq:initialcurvature} for any support submanifold $\tilde{P}$ of $P$ and any future pointing normal null geodesic staring from $P$
\end{enumerate}
Then $E^+(P) \cap P$ is achronal, and $E^+(E^+(P) \cap P)$ is compact or $M$ is null geodesically incomplete.
\end{proposition}

Finally we turn to (I4) and introduce trapped points in the current setting. In \cite[Sec.\ 6.3]{GGKS:18}, a faithful generalisation of the classical condition to the $C^{1,1}$-setting is given, motivated by Jacobi tensor classes and the mean curvature of spacelike $2$-surfaces, given as the level sets of the exponential map that generate the light cone. Although these tools are no longer at our disposal we can use the very formulation which, once more, is given in the support sense:

\begin{definition}[Trapped points in $C^1$]\label{definition: trappedset}
A point $p \in M$ is called \textit{future trapped}, if for any future pointing null vector $v \in T_pM$ and for any null geodesic $\gamma$ with $\gamma(0) = p$, $\dot{\gamma}(0)=v$, there exists a parameter $t$ and a spacelike $C^2$-submanifold $\Tilde S$ of codimension $m=2$ with $\Tilde{S}\subseteq J^+(p)$, $\gamma(t) \in \Tilde{S}$ and convergence ${k}_{\Tilde{S}}(\dot{\gamma}(t)) > 0$. 
\end{definition}

Now the techniques established above allow one to also prove (\cite[Prop.\ 5.13]{KOSS:22}) that in any strongly causal, null geodesically complete, MNNB $C^1$-spacetime satisfying \eqref{eq:dnec} the horismos $E^+(p)$ is compact for any future trapped point $p$, and we have established (C3) once more.
\medskip

Now we are finally in a position to put everything together and establish the main result in our account.

\subsection{The Hawking-Penrose theorem in $C^1$}\label{sec:4.10}

Finally, we are in the position to formulate the long seeked result, namely the analytical Hawking-Penrose theorem for maximally causally non-branching $C^1$-spacetimes. We first give the precise statement and comment on some of its specifics, and then put together the arguments that allow to derive it from the causal result given in Lemma \ref{thm:C1HPcausalityversion}. 

Indeed, the following result that appeared as Theorem 6.3 in \cite{KOSS:22} generalises  the classical Theorem \ref{thm:CHP} as well as the $C^{1,1}$-version given in \cite[Thms.\ 2.6, 2.6]{GGKS:18}.

\pagebreak%artificial

\begin{theorem}[$C^1$-Hawking-Penrose]\label{Theorem: HPC1}
Let $(M,g)$ be a $C^1$-spacetime such that 
\begin{itemize}
    \item[(E)] the distributional strong energy condition \eqref{eq:dsec} holds as well as the distributional genericity condition \eqref{eq:dgc} along any causal geodesic,
    \item[(C)] it is causal, and
    \item[(B)] it is maximally causally non-branching (MCNB).
\end{itemize}
Moreover, assume it contains at least one of the following:
\begin{enumerate}
\item[(I1)] a compact achronal set without edge,
\item[(I2)] a closed future trapped $C^0$-surface $P$,
\item[(I3)] a closed future trapped $C^0$-submanifold $P$ of co-dimension $2 < m < n$ such that 
\eqref{eq:initialcurvature} holds for any support submanifold and any future pointing normal null geodesic staring from it, 
\item[(I4)] a future trapped point in the sense of Definition \ref{definition: trappedset}.
\end{enumerate}
Then $M$ is causally geodesically incomplete.
\end{theorem}

We now briefly compare the present result with its classical counterpart. The distributional energy conditions used in (E) are faithful generalisations of the classical conditions. The causality condition (C) is slightly stronger than the condition of chronology used in the classical result and we have commented on our use of causality at the end of Section \ref{sec:4.8}. At the moment it is not clear whether this is a mere technical point or whether causality is strictly necessary for the result to hold. The initial conditions (I1)--(I4) are again faithful generalisations of their classical counterparts and their formulation in the support sense seems only natural in the $C^1$-regularity class. 

Finally, we come to condition (B), which is entirely new. As discussed in Section \ref{sec:4.8} it is, on the one hand, necessary to secure a main argument in the proof of the the advanced focusing result \eqref{nb-motivation}, which is the fundamental ingredient to establish the inexistence of lines under \eqref{eq:dsec}/\eqref{eq:dnec} and \eqref{eq:dgc}. On the other hand, it is well motivated by similar conditions used in metric geometry, see also Section \ref{sec:5}, below. Moreover, it adds a novel aspect to the interpretation of the $C^1$-theorem: Under the given conditions the result predicts either geodesic incompleteness or branching of maximising causal geodesics. The latter alternative physically signifies an event equally catastrophic for the corresponding observer or light ray: instead of suddenly beginning or ending its existence it splits in two or, in the past case, two observers/light rays are merged into one.

Of course, there is still the alternative that the regularity of the metric drops even further, i.e., below $C^1$, which renders the curvature a distribution of higher order. Again, the result also forbids the extension of the spacetime to a complete one of regularity $C^1$ without (maximal causal) geodesic branching and we shall briefly return to this discussion in the concluding Section \ref{sec:5}.
\medskip

\noindent\emph{Sektch of proof.}
Assuming causal geodesic completeness, we once again deduce a contradiction, now using Lemma \ref{thm:C1HPcausalityversion}.
%\medskip

First note that (C) implies (C1) there.
Then Theorem \ref{Theorem:nolines} implies the assumptions (C2') and (C2'') of Theorem \ref{thm:C1HPcausalityversion} and it remains to establish the existence of a trapped set, i.e.\ (C3) for which we have to distinguish the four different cases.

The case for (I1) has been made at the beginning of Section \ref{sec:4.8}. 
For the remaining cases we need---just as in the classical proof of Theorem \ref{thm:CHP}---an ``upgrade'' of (C) to strong causality. In the present setting this can be achieved from the fact that there are no null lines, cf.\ \cite[Lem.\ A.31]{KOSS:22}. Then Proposition \ref{Proposition:submfdnullcompletenessortrappedset} covers the cases (I2) and (I3), while we have argued for case (I4) already at the end of Section \ref{sec:4.8}.
\hfill\qed
\medskip

With these arguments we have finished our account on the extension of all three of the major singularity theorems of GR to regularity $C^1$ and we proceed to some conclusions and a general discussion. 

\section{Conclusions, alternatives \& outlook}\label{sec:5}

We begin this final section with a summary and some conclusions drawn from the results presented here. Then we will discuss some further perspectives of the approach at hand as well as alternative approaches to singularity theorems beyond the smooth setting.
\medskip

In this review we have discussed the extension to $C^1$-spacetimes of the classical singularity theorems of GR, which under physically reasonable condition assert causal geodesic incompleteness. These results, as well as the $C^1$-extension of the Gannon-Lee theorem in \cite{SS:21}, which we have avoided to discuss here, are entirely in the spirit of the classical theorems. That is, they extend the original conditions in a natural way, use essentially the same line of arguments and come to the same conclusions as the classical results, however, with one noteworthy extension: The Hawking-Penrose and the Gannon-Lee theorem add a further alternative to causal geodesic incompleteness, namely the branching of maximising causal geodesics. 

The proofs are, at the one hand, based on the recent extensions of causality theory to low regularity Lorentzian metrics and, on the other hand, rely on an extension of the focusing results for causal geodesics. Indeed the latter, which is achieved via a regularisation approach, is the main technical advance presented here.
%, which is hence more on the analytical side of the arguments.
\medskip

Next we briefly discuss the further prospects of this approach. It is generally expected that the causality parts of the results extend to $C^{0,1}$-metrics,\footnote{A class, for which Hawking and Ellis \cite[p.\ 268f]{HE:73} still speculate the singularity theorems to hold, cf.\ also Section \ref{sec:4.1}.} since such metrics still belong to the so-called \emph{causally plain} ones, cf.\ \cite[Def.\ 1.16]{CG:12}. The latter allow for essentially the same causality theory as smooth spacetimes, see \cite[Thm.\ 1.25]{CG:12}. Indeed, it is only below Lipschitz regularity that such core features of causality theory as the push-up principle and the openness of $I^+$ become an issue \cite{GKS:19}, and the lightcones may form subsets of full measure \cite[Ex.\ 1.11]{CG:12}. 

On the analytic side, it seems also feasible to extend the recent techniques to locally Lipschitz metrics. 
Here, one primary task is to extend the Friedrichs Lemma \ref{lem:Fthesecond}, which  lies at the analytical core of the regularisation techniques, as it allows to derive from the distributional energy conditions of the singular metric useful surrogate energy conditions for the smooth approximations. Next, Lemma \ref{Lemma: approximatingmetrics} which gives the precise speed of convergence of the regularised metrics to the singular one would have to be revisited. Also, the formulation of the energy conditions becomes more subtle since the curvature ceases to be an order-one distribution which forecloses the insertion of $C^1$-vector fields which was  essential at least for the genericity condition. Finally, one faces the problem that the right hand side of the geodesic equation now is merely locally bounded and it seems unavoidable to resort to non-classical solution concepts such as Filippov solutions \cite{Fil:88}, see also the discussion in Section \ref{sec:4.1}.

Still this is actually a long way from the largest possible class where the curvature can be (stably) defined in an analytical (distributional) way, i.e.\ the GT-class $H^2_{\mbox{{\tiny loc}}}\cap L^\infty_{\mbox{{\tiny loc}}}$, see Section \ref{sec:4.2}. However, as remarked there, the quest is to at least go into the direction of regularity classes more closely linked to the PDE-approach to GR, e.g.\ $g\in H^{5/2+\eps}$ or $\nabla\in L^2$. 
\medskip

A somewhat related topic is singularities in semi-classical and quantum theories of gravity. There the 
energy conditions are expected to be violated and to hold only in some averaged sense, see e.g.\ \cite{Vis:95}. The question then arises whether under such assumptions the focusing effect persists, 
or whether the singularity theorems vanish altogether in the quantum regime, see \cite[Sec.\ 8.2]{SG:15}, \cite[Sec.\ 6.2]{Sen:98}. Indeed, focusing can be maintained under energy conditions averaged along causal geodesics, see e.g.\ \cite{FG:11} and the references therein. Similar results using index form techniques have appeared in \cite{FK:20} and recent work is concerned with energy conditions directly related to quantum energy inequalities \cite{FFK:21} and worldvolume energy inequalities \cite{GKOS:22}. There is certainly a technical proximity of the methods used 
there and the ones described in this review, and future research will investigate their interrelations more closely.
\medskip

Now shifting away from the more analytic parts of the singularity theorems we first turn to causality theory. It has long been clear that causality theory is very robust in general. Most arguments are rather topological in nature and can actually be seen as belonging to frameworks more abstract than Lorentzian differential geometry. Indeed, Ettore Minguzzi in \cite{Min:19} has recently put forward a very general theory of \emph{causal cone structures}, that is a theory of upper semi-continuous distributions of cones over manifolds (which generalise the lightcone of Lorentzian metrics). In this setting it is indeed possible to establish the causal core of some of the singularity theorems: One may see the analytic concepts like the energy conditions, focusing results, etc.\ used throughout this note merely as tools that produce subsets in $M$ that possess specific causality properties. Completely removing them from the arguments one arrives at purely causal results. To give some flavour of these, we quote a version of the Penrose theorem which appeared as \cite[Thm.\ 2.67]{Min:19}, for more details see Sec.\ 2.15 there.

\begin{theorem}[Causal Penrose Theorem]
 Let $(M,C)$ be a globally  hyperbolic closed  cone  structure  admitting  a  non-compact  stable  Cauchy hypersurface. Then there are no compact future trapped sets and if $S$ is non-empty and compact there is an inextendible future null geodesic entirely contained in $E^+(S)$.
\end{theorem}

Another and quite different approach to singularity theorems has been opened up in the context of the recently developed synthetic approach to Lorentzian geometry put forward by Michael Kunzinger and Clemens S\"amann in \cite{KS:18}. The \emph{Lorentzian length spaces} introduced there are the analogue of metric length spaces, which have long been used as an essential tool to extract the metric core of many notions and results in Riemannian geometry, see e.g.\ \cite{BBI:01}. Lorentzian pre-length spaces $(X,d,\leq,\ll,\tau)$ are metric spaces $(X,d)$ together with with a preorder $\leq$ and  a transitive relation $\ll$ contained in $\leq$ (which model the causal and timelike relations of Lorentzian geometry) and a lower semi-continuous map $\tau:X\times X \to [0,\infty]$ that satisfies the reverse triangle inequality (and models the Lorentzian distance function). Such a space is called a Lorentzian length space, if, in addition to some technical conditions,  $\tau$ is intrinsic in the sense that the distance between points defined via the $\sup$ of the $\tau$-length of connecting causal curves coincides with their $\tau$-distance. Causality theory in Lorentzian length spaces \cite{KS:18,GKS:19,ACS:20,BORS:22} extends standard causality theory beyond the spacetime setting to which it reduces for continuous spacetimes with strongly causal and causally plain metric.  

In this setting it becomes possible to formulate synthetic versions of the singularity theorems, in the sense that the energy conditions are implemented as synthetic curvature bounds. A first theorem in Lorentzian length spaces that are warped products and which uses suitable sectional curvature bounds (implying Ricci curvature bounds in such geometries), based on triangle comparison is the following version of the Hawking theorem, cf.\ \cite[Cor 6.2(ii)]{AGKS:22}:

\begin{theorem}[Synthetic Hawking Theorem] Let $X$ be a geodesic length space,\footnote{That is a metric length space where each pair of points can be joined by a minimising curve.} and let $Y=I\times_f X$ be a warped procduct with $I=(a,b)$ and $f:I\to (0,\infty)$ smooth.\footnote{In the space $Y$, $\tau$ is defined via the $\sup$ of the length of future directed causal curves $\gamma=(\alpha,\beta)$, given by $L(\gamma)=\int\sqrt{\dot\alpha^2-(f\circ\alpha)^2v_\beta^2}$ with $v_\beta$ the metric derivative. Observe that this precisely models the smooth situation.}. Assume that $Y$ has timelike sectional curvature bounded below by $0$ and that $f$ is non-constant. Then  $a>-\infty $ or $b<\infty$ and hence $Y$ is past or future timelike geodesically incomplete.
\end{theorem}

Of course, implementing the classical energy conditions \eqref{eq:sec} or \eqref{eq:nec} rather amounts to Ricci curvature bounds than to sectional ones. Indeed, synthetic Ricci curvature bounds have been intensively studied in Riemannian geometry using optimal transport, see e.g.\ \cite{Vil:09}. These techniques have recently been transferred to the \emph{smooth} Lorentzian setting setting in \cite{MCC:20,MS:22} and further extended to
the synthetic setting of Lorentzian length spaces by Fabio Cavaletti and Andrea Mondino in \cite{CM:20}.
The basic idea is that timelike lower Ricci bounds can be characterised in terms of the convexity of an entropy functional along $l_p$-geodesics in the space of probability measures, where $l_p$ is the Lorentz-Wasserstein distance. The corresponding timelike curvature-dimension conditions TCD(K,N) and its weaker variant the timelike measure contraction property TMCP(K,N) then allow to formulate a version of the Hawking singularity theorem which we here quote in a loose way omitting technicalities, cf.\ \cite[Thm.\ 5.2]{CM:20} for the precise version:

\pagebreak%artificial

\begin{theorem}[TMCP-Hawking Theorem] Let $X$ be a timelike non-branching, globally hyperbolic Lorentzian pre-length space satisfying a TMCP-property. Let $V$ be a Borel achronal future timelike complete subset with mean curvature bounded above. Then every future timelike geodesic starting in $V$ has a bounded maximal domain of existence. 
\end{theorem}
\medskip

To sum up, in this review we have discussed the classical singularity theorems of GR and sketched the main arguments leading to their proofs. One may actually identify two main lines in these arguments, the analytical and the causal one. The former is concerned with providing focusing results for causal geodesics using the energy conditions which lead to the occurrence of conjugate or focal points and hence provide estimates on when causal geodesics stop maximising the Lorentzian length. The second, causal line of arguments 
%which originates from the study of the future and past of points and sets in spacetime 
gives criteria for maximising causal geodesic to exist. Confronting these two threads leads to a contradiction unless some of the causal geodesics become incomplete.

In the main part of this work we have extended the causal and the analytic line of arguments to Lorentzian metrics of regularity $C^1$ and have presented corresponding extensions of the classical theorems. The main achievements presented are on the analytic side using a regularisation approach that allows to deal with the distributional curvature associated with a $C^1$-metric. 

In the final discussion we have complemented these results with an overview of recent versions of singularity theorems in a purely causal setting as well as in a synthetic setting. It is clear that these results and techniques are still fresh, and that many interesting lines of research emerge from here. Also, it is unclear to date how the synthetic results precisely relate to the analytical approach put forward here and its possible extension to even lower regularity as discussed above, but see \cite{KOV:22} for a first work addressing this issue. 
\medskip

In any case, it can firmly be stated that the singularity theorems are not only an integral part of GR and Lorentzian geometry but, even more than half a century after they first emerged, they are still an interesting field of research holding many quests to be resolved in the future---both from the physical side, see e.g.\ \cite[Sec.\ 8]{SG:15}, as well as from a mathematical perspective as laid out here.

\section*{Acknowledgement}
The author wishes to thank his frequent collaborators Michael Kunzinger, Clemens Sämann, James Vickers, and James Grant for their friendship and support as well as our joint (former) students Milena Stojkovi\'c, Melanie Graf, Benedict Schinnerl, and Argam Ohanyan, who have contributed so much to the whole enterprise. This work was supported by FWF-project P33594 of the Austrian Science Fund.

%\bibliography{ro}
%\end{document}

%  \bibliographystyle{alpha}
% \bibliography{ro}

\begin{thebibliography}{AHCPS20}

\bibitem[Aea16]{GW:15}
B.~P. Abbott and et~al.
\newblock Observation of gravitational waves from a binary black hole merger.
\newblock {\em Phys. Rev. Lett.}, 116(6):061102, 16, 2016.
\newblock Authors include B. C. Barish, K. S. Thorne and R. Weiss.

\bibitem[AGKC22]{AGKS:22}
Stephanie~B. Alexander, Melanie Graf, Michael Kunzinger and S\"amann Clemens.
\newblock Generalized cones as lorentzian length spaces: Causality, curvature,
  and singularity theorems.
\newblock {\em Comm. Anal. Geom.}, to appear., 2022.

\bibitem[AHCPS20]{ACS:20}
Luis Ak\'{e}~Hau, Armando~J. Cabrera~Pacheco and Didier~A. Solis.
\newblock On the causal hierarchy of {L}orentzian length spaces.
\newblock {\em Classical Quantum Gravity}, 37(21):215013, 22, 2020.

\bibitem[BS10]{BS:10}
Thomas~W. Baumgarte and Stuart~L. Shapiro.
\newblock {\em Numerical Relativity: Solving Einstein's Equations on the
  Computer}.
\newblock Cambridge University Press, Cambridge, 2010.

\bibitem[BORS22]{BORS:22}
Tobias Beran, Argam Ohanyan, Felix Rott and Didier Solis.
\newblock The splitting theorem for globally hyperbolic lorentzian length spaces with
  non-negative timelike curvature.
\newblock {\em arXiv 2209.14724 [math.DG]}, 2022.

\bibitem[BS07]{BS:07}
Antonio~N. Bernal and Miguel S{\'a}nchez.
\newblock Globally hyperbolic spacetimes can be defined as `causal' instead of
  `strongly causal'.
\newblock {\em Classical Quantum Gravity}, 24(3):745--749, 2007.

\bibitem[BS18]{BS:18}
Patrick Bernard and Stefan Suhr.
\newblock Lyapounov {F}unctions of {C}losed {C}one {F}ields: {F}rom {C}onley
  {T}heory to {T}ime {F}unctions.
\newblock {\em Comm. Math. Phys.}, 359(2):467--498, 2018.

\bibitem[BEE96]{BEE:96}
John~K. Beem, Paul~E. Ehrlich and Kevin~L. Easley.
\newblock {\em Global {L}orentzian geometry}, volume 202 of {\em Monographs and
  Textbooks in Pure and Applied Mathematics}.
\newblock Marcel Dekker Inc., New York, second edition, 1996.

\bibitem[BGY17]{BGN:17}
Lydia Bieri, David Garfinkle and Nicol\'{a}s Yunes.
\newblock Gravitational waves and their mathematics.
\newblock {\em Notices Amer. Math. Soc.}, 64(7):693--707, 2017.

\bibitem[BBI01]{BBI:01}
Dmitri Burago, Yuri Burago and Sergei Ivanov.
\newblock {\em A course in metric geometry}, volume~33 of {\em Graduate Studies
  in Mathematics}.
\newblock American Mathematical Society, Providence, RI, 2001.

\bibitem[CG12]{CG:12}
Piotr~T. Chru{\'s}ciel and James D.~E. Grant.
\newblock On {L}orentzian causality with continuous metrics.
\newblock {\em Classical Quantum Gravity}, 29(14):145001, 32, 2012.

\bibitem[Chr09]{Chr:09}
Demetrios Christodoulou.
\newblock {\em The formation of black holes in general relativity}.
\newblock EMS Monographs in Mathematics. European Mathematical Society (EMS),
  Z\"{u}rich, 2009.

\bibitem[Chr11]{Chr:11}
Piotr~T. Chru{\'s}ciel.
\newblock Elements of causality theory.
\newblock 2011.
\newblock arXiv:1110.6706 [gr-qc].

\bibitem[Cla93]{Cla:93}
Chris J.~S. Clarke.
\newblock {\em The analysis of space-time singularities}, volume~1 of {\em
  Cambridge Lecture Notes in Physics}.
\newblock Cambridge University Press, Cambridge, 1993.

\bibitem[CM20]{CM:20}
Fabio Cavalletti and Andrea Mondino.
\newblock Optimal transport in lorentzian synthetic spaces, synthetic timelike
  ricci curvature lower bounds and applications.
\newblock 2020.

\bibitem[Daf21]{Daf:21}
Mihalis Dafermos.
\newblock Penrose's incompleteness theorem.
\newblock {\em LMS Newsletter}, 493:27--34, 2021.

\bibitem[EH90]{EH:90}
Jost-Hinrich Eschenburg and Ernst Heintze.
\newblock Comparison theory for {R}iccati equations.
\newblock {\em Manuscripta Math.}, 68(2):209--214, 1990.

\bibitem[FB52]{CB:52}
Yvonne. Four\`es-Bruhat.
\newblock Th\'{e}or\`eme d'existence pour certains syst\`emes d'\'{e}quations
  aux d\'{e}riv\'{e}es partielles non lin\'{e}aires.
\newblock {\em Acta Math.}, 88:141--225, 1952.

\bibitem[FFK21]{FFK:21}
Jackson~R. Fliss, Ben Freivogel and Eleni-Alexandra Kontou.
\newblock The double smeared null energy condition.
\newblock {\em ArXiv 2111.05772}, 2021.

\bibitem[FG11]{FG:11}
Christopher~J. Fewster and Gregory~J. Galloway.
\newblock Singularity theorems from weakened energy conditions.
\newblock {\em Classical Quantum Gravity}, 28(12):125009, 18, 2011.

\bibitem[Fil88]{Fil:88}
Aleksei~F. Filippov.
\newblock {\em Differential equations with discontinuous righthand sides},
  volume~18 of {\em Mathematics and its Applications (Soviet Series)}.
\newblock Kluwer Academic Publishers Group, Dordrecht, 1988.

\bibitem[FK20]{FK:20}
Christopher~J. Fewster and Eleni-Alexandra Kontou.
\newblock A new derivation of singularity theorems with weakened energy
  hypotheses.
\newblock {\em Classical Quantum Gravity}, 37(6):065010, 31, 2020.

\bibitem[FS12]{FS:12}
Albert Fathi and Antonio Siconolfi.
\newblock On smooth time functions.
\newblock {\em Math. Proc. Cambridge Philos. Soc.}, 152(2):303--339, 2012.

\bibitem[Gan75]{Gan:75}
Dennis Gannon.
\newblock Singularities in nonsimply connected space-times.
\newblock {\em J. Mathematical Phys.}, 16(12):2364--2367, 1975.

\bibitem[GGKS18]{GGKS:18}
Melanie Graf, James D.~E. Grant, Michael Kunzinger and Roland Steinbauer.
\newblock The {H}awking--{P}enrose {S}ingularity {T}heorem for
  {$C^{1,1}$}-{L}orentzian {M}etrics.
\newblock {\em Comm. Math. Phys.}, 360(3):1009--1042, 2018.

\bibitem[GKOS01]{GKOS:01}
Michael Grosser, Michael Kunzinger, Michael Oberguggenberger and Roland
  Steinbauer.
\newblock {\em Geometric theory of generalized functions with applications to
  general relativity}, volume 537 of {\em Mathematics and its Applications}.
\newblock Kluwer Academic Publishers, Dordrecht, 2001.

\bibitem[GKS19]{GKS:19}
James D.~E. Grant, Michael Kunzinger and Clemens S\"{a}mann.
\newblock Inextendibility of spacetimes and {L}orentzian length spaces.
\newblock {\em Ann. Global Anal. Geom.}, 55(1):133--147, 2019.

\bibitem[GKSS20]{GKSS:20}
James D.~E. Grant, Michael Kunzinger, Clemens S\"{a}mann and Roland
  Steinbauer.
\newblock The future is not always open.
\newblock {\em Lett. Math. Phys.}, 110(1):83--103, 2020.

\bibitem[GLS18]{GLS:18}
Gregory~J. Galloway, Eric Ling and Jan Sbierski.
\newblock Timelike completeness as an obstruction to {$C^0$}-extensions.
\newblock {\em Comm. Math. Phys.}, 359(3):937--949, 2018.

\bibitem[GP09]{GP:09}
Jerry~B. Griffiths and Ji\v{r}\'{\i} Podolsk\'y.
\newblock {\em Exact space-times in {E}instein's general relativity}.
\newblock Cambridge Monographs on Mathematical Physics. Cambridge University
  Press, Cambridge, 2009.

\bibitem[Gra20]{G:20}
Melanie Graf.
\newblock Singularity theorems for {$C^1$}-{L}orentzian metrics.
\newblock {\em Comm. Math. Phys.}, 378(2):1417--1450, 2020.

\bibitem[GKOS22]{GKOS:22}
Melanie Graf,  Eleni-Alexandra Kontou, Argam Ohanyan and Benedict Schinnerl.
\newblock Hawking-type singularity theorems for worldvolume energy inequalities.
\newblock {\em arXiv 2209.04347 [gr-qc]}, 2022.


\bibitem[GS10]{GS:10}
Gregory~J. Galloway and Jos\'{e} M.~M. Senovilla.
\newblock Singularity theorems based on trapped submanifolds of arbitrary
  co-dimension.
\newblock {\em Classical Quantum Gravity}, 27(15):152002, 10, 2010.

\bibitem[G68]{G:68}
Robert Geroch.
\newblock {What is a singularity in general relativity?}
\newblock {\em Ann.\ Phys.}, {48}(3), {526}--{540}, 1968.

\bibitem[GT87]{GT:87}
Robert Geroch and Jennie Traschen.
\newblock Strings and other distributional sources in general relativity.
\newblock {\em Phys.~Rev.~D}, {36}(4):1017--1031, 1987.

\bibitem[H\"03]{Hoe:03}
Lars H\"{o}rmander.
\newblock {\em The analysis of linear partial differential operators. {I}}.
\newblock Classics in Mathematics. Springer-Verlag, Berlin, 2003.
\newblock Distribution theory and Fourier analysis, Reprint of the second
  (1990) edition [Springer, Berlin].

\bibitem[Har02]{Har:02}
Philip Hartman.
\newblock {\em Ordinary differential equations}, volume~38 of {\em Classics in
  Applied Mathematics}.
\newblock Society for Industrial and Applied Mathematics (SIAM), Philadelphia,
  Pa., 2002.
\newblock Corrected reprint of the second (1982) edition [Birkh\"{a}user,
  Boston, MA; MR0658490 (83e:34002)], With a foreword by Peter Bates.

\bibitem[Haw67]{Haw:67}
Stephen~W. Hawking.
\newblock The occurrence of singularities in cosmology. iii. causality and
  singularities.
\newblock {\em Proceedings of the Royal Society of London. Series A,
  Mathematical and Physical Sciences}, 300(1461):187--201, 1967.

\bibitem[HE73]{HE:73}
Stephen~W. Hawking and George F.~R. Ellis.
\newblock {\em The large scale structure of space-time}.
\newblock Cambridge University Press, London, New York, 1973.
\newblock Cambridge Monographs on Mathematical Physics, No. 1.

\bibitem[HM19]{HM:19}
Raymond~A. Hounnonkpe and Ettore Minguzzi.
\newblock Globally hyperbolic spacetimes can be defined without the `causal'
  condition.
\newblock {\em Classical Quantum Gravity}, 36(19):197001, 9, 2019.

\bibitem[HP70]{HP:70}
Stephen~W. Hawking and Roger Penrose.
\newblock The singularities of gravitational collapse and cosmology.
\newblock {\em Proc. Roy. Soc. London Ser. A}, 314:529--548, 1970.

\bibitem[HS02]{HS:02}
J.~Mark Heinzle and Roland Steinbauer.
\newblock Remarks on the distributional {S}chwarzschild geometry.
\newblock {\em J. Math. Phys.}, 43(3):1493--1508, 2002.

\bibitem[HW51]{HW:51}
Philip Hartman and Aurel Wintner.
\newblock On the problems of geodesics in the small.
\newblock {\em Amer. J. Math.}, 73:132--148, 1951.

\bibitem[Isr66]{Isr:66}
Werner Israel.
\newblock Singular hypersurfaces and thin shells in general relativity.
\newblock {\em Il Nuovo Cimento B}, 44(1):1--14, 1966.

\bibitem[KOSS22]{KOSS:22}
Michael Kunzinger, Argam Ohanyan, Benedict Schinnerl and Roland Steinbauer.
\newblock The {H}awking--{P}enrose {S}ingularity {T}heorem for
  {$C^{1}$}-{L}orentzian {M}etrics.
\newblock {\em Comm. Math. Phys.}, (to appear), 2022.

\bibitem[Kri99]{Kri:99}
Marcus Kriele.
\newblock {\em Spacetime}, volume~59 of {\em Lecture Notes in Physics. New
  Series m: Monographs}.
\newblock Springer, Berlin, 1999.
\newblock Foundations of general relativity and differential geometry.

\bibitem[KS18]{KS:18}
Michael Kunzinger and Clemens S\"{a}mann.
\newblock Lorentzian length spaces.
\newblock {\em Ann. Global Anal. Geom.}, 54(3):399--447, 2018.

\bibitem[KSS14]{KSS:14}
Michael Kunzinger, Roland Steinbauer and Milena Stojkovi{\'c}.
\newblock The exponential map of a {$C^{1,1}$}-metric.
\newblock {\em Differential Geom. Appl.}, 34:14--24, 2014.

\bibitem[KSSV14]{KSSV:14}
Michael Kunzinger, Roland Steinbauer, Milena Stojkovi{\'c} and James~A.
  Vickers.
\newblock A regularisation approach to causality theory for
  {$C^{1,1}$}-{L}orentzian metrics.
\newblock {\em Gen. Relativity Gravitation}, 46(8):Art. 1738, 18, 2014.

\bibitem[KSSV15]{KSSV:15}
Michael Kunzinger, Roland Steinbauer, Milena Stojkovi{\'c} and James~A.
  Vickers.
\newblock Hawking's singularity theorem for {$C^{1,1}$}-metrics.
\newblock {\em Classical Quantum Gravity}, 32(7):075012, 19, 2015.

\bibitem[KSV15]{KSV:15}
Michael Kunzinger, Roland Steinbauer and James~A. Vickers.
\newblock The {P}enrose singularity theorem in regularity {$C^{1,1}$}.
\newblock {\em Classical Quantum Gravity}, 32(15):155010, 12, 2015.

\bibitem[LAC22]{LLS:22}
Christian Lange, Lytchak Alexander and S{\"a}mann Clemens.
\newblock Lorentz meets lipschitz.
\newblock {\em Adv. Theor. Math. Phys.}, to appear, 2022.

\bibitem[KOV22]{KOV:22}
Michael Kunzinger, Michael Oberguggenberger and James A. Vickers.
\newblock Synthetic versus distributional lower ricci curvature bounds.
\newblock {\em arXiv 2207.03715 [math.DG]}, 2022.

\bibitem[Lan21]{Lan:21}
Klaas Landsman.
\newblock Singularities, black holes and cosmic censorship: a tribute to
  {R}oger {P}enrose.
\newblock {\em Found. Phys.}, 51(2):Paper No. 42, 38, 2021.
\newblock With an appendix by Erik Curiel.

\bibitem[Lee76]{Lee:76}
Charles~W. Lee.
\newblock A restriction on the topology of {C}auchy surfaces in general
  relativity.
\newblock {\em Comm. Math. Phys.}, 51(2):157--162, 1976.

\bibitem[Lic55]{Lic:55}
Andr\'{e} Lichnerowicz.
\newblock {\em Th\'eories relativistes de la gravitation et de
  l'\'electromagn\'etisme. {R}elativit\'e g\'en\'erale et th\'eories
  unitaires}.
\newblock Masson et Cie, Paris, 1955.

\bibitem[LM07]{LM:07}
Philippe~G. LeFloch and Cristinel Mardare.
\newblock Definition and stability of {L}orentzian manifolds with
  distributional curvature.
\newblock {\em Port. Math. (N.S.)}, 64(4):535--573, 2007.

\bibitem[Mar68]{Mar:68}
Jerrold~E. Marsden.
\newblock Generalized {H}amiltonian mechanics: {A} mathematical exposition of
  non-smooth dynamical systems and classical {H}amiltonian mechanics.
\newblock {\em Arch. Rational Mech. Anal.}, 28:323--361, 1967/68.

\bibitem[MAS15]{MS:15}
Pablo Morales~\'{A}lvarez and Miguel S\'{a}nchez.
\newblock Myers and {H}awking theorems: geometry for the limits of the
  universe.
\newblock {\em Milan J. Math.}, 83(2):295--311, 2015.

\bibitem[McC20]{MCC:20}
Robert~J. McCann.
\newblock Displacement convexity of {B}oltzmann's entropy characterizes the
  strong energy condition from general relativity.
\newblock {\em Camb. J. Math.}, 8(3):609--681, 2020.

\bibitem[Min09]{Min:09a}
Ettore Minguzzi.
\newblock Characterization of some causality conditions through the continuity
  of the {L}orentzian distance.
\newblock {\em J. Geom. Phys.}, 59(7):827--833, 2009.

\bibitem[Min15]{Min:15}
Ettore Minguzzi.
\newblock Convex neighborhoods for {L}ipschitz connections and sprays.
\newblock {\em Monatsh. Math.}, 177(4):569--625, 2015.

\bibitem[Min19a]{Min:19}
Ettore Minguzzi.
\newblock Causality theory for closed cone structures with applications.
\newblock {\em Rev. Math. Phys.}, 31(5):1930001, 139, 2019.

\bibitem[Min19b]{Min:19a}
Ettore Minguzzi.
\newblock Lorentzian causality theory.
\newblock {\em Living Rev.\ Relativ.}, 22(3):220 pp, 2019.

\bibitem[MS93]{MS:93}
Marc Mars and Jos{\'e} M.~M. Senovilla.
\newblock Geometry of general hypersurfaces in spacetime: junction conditions.
\newblock {\em Classical Quantum Gravity}, 10(9):1865--1897, 1993.

\bibitem[MS08]{MS:08}
Ettore Minguzzi and Miguel S{\'a}nchez.
\newblock The causal hierarchy of spacetimes.
\newblock In {\em Recent developments in pseudo-{R}iemannian geometry}, ESI
  Lect. Math. Phys., pages 299--358. Eur. Math. Soc., Z\"urich, 2008.

\bibitem[MS22]{MS:22}
Andrea Mondino and Stefan Suhr.
\newblock An optimal transport formulation of the einstein equations of general
  relativity.
\newblock {\em Journal of the European Mathematical Society, to appear}, 2022.

\bibitem[MTW73]{MTW:73}
Charles~W. Misner, Kip~S. Thorne and John~Archibald Wheeler.
\newblock {\em Gravitation}.
\newblock W. H. Freeman and Co., San Francisco, Ca., 1973.

\bibitem[Nig13]{Nig:13}
Eduard~A. Nigsch.
\newblock Bornologically isomorphic representations of distributions on
  manifolds.
\newblock {\em Monatsh.\ Math.} {170}(1), {49}--{63}, 2013.


\bibitem[Oha22]{Oha:22}
Argam Ohanyan.
\newblock Geometric foundations of the classical singularity theorems.
\newblock Master's thesis, University of Vienna, 2022.

\bibitem[O'N83]{ONe:83}
Barrett O'Neill.
\newblock {\em Semi-{R}iemannian geometry with applications to relativity},
  volume 103 of {\em Pure and Applied Mathematics}.
\newblock Academic Press, New York, 1983.

\bibitem[OS39]{OS:39}
Robert~J. Oppenheimer and Hartland Snyder.
\newblock On continued gravitational contraction.
\newblock {\em Phys.\ Rev. (2)}, 56(5):455--459, 1939.

\bibitem[Pen65]{Pen:65}
Roger Penrose.
\newblock Gravitational collapse and space-time singularities.
\newblock {\em Phys. Rev. Lett.}, 14:57--59, 1965.

\bibitem[Pen72]{Pen:72b}
Roger Penrose.
\newblock {\em Techniques of differential topology in relativity}.
\newblock Society for Industrial and Applied Mathematics, Philadelphia, Pa.,
  1972.
\newblock Conference Board of the Mathematical Sciences Regional Conference
  Series in Applied Mathematics, No. 7.

\bibitem[Rin15]{Rin:15}
Hans Ringstr\"{o}m.
\newblock Origins and development of the {C}auchy problem in general
  relativity.
\newblock {\em Classical Quantum Gravity}, 32(12):124003, 37, 2015.

\bibitem[S{\"a}m16]{Sae:16}
Clemens S{\"a}mann.
\newblock Global hyperbolicity for spacetimes with continuous metrics.
\newblock {\em Ann. Henri Poincar\'e}, 17(6):1429--1455, 2016.

\bibitem[Sbi18]{Sbi:18}
Jan Sbierski.
\newblock The {$C^0$}-inextendibility of the {S}chwarzschild spacetime and the
  spacelike diameter in {L}orentzian geometry.
\newblock {\em J. Differential Geom.}, 108(2):319--378, 2018.

\bibitem[Sen98]{Sen:98}
Jos\'e M.~M. Senovilla.
\newblock Singularity theorems and their consequences.
\newblock {\em Gen. Relativity Gravitation}, 30(5):701--848, 1998.

\bibitem[Sen12]{Sen:12}
Jos\'e M.~M. Senovilla.
\newblock Singularity theorems in general relativity: Achievements and open
  questions.
\newblock In Christoph Lehner, Jürgen Renn and Schemmel Matthias, editors,
  {\em Einstein and the Changing Worldviews of Physics}, pages 305--315.
  Birkh\"auser, New York, 2012.

\bibitem[Sen21]{Sen:21}
Jos\'e M.~M. Senovilla.
\newblock A critical appraisal of the singularity theorems.
\newblock {\em arXiv:2108.07296 [gr-qc]}, 2021.

\bibitem[SG15]{SG:15}
Jos\'e M.~M. Senovilla and David Garfinkle.
\newblock The 1965 {P}enrose singularity theorem.
\newblock {\em Classical Quantum Gravity}, 32(12):124008, 45, 2015.

% \bibitem[Sh99]{Sta:99}
% Fredrik St\aa~hl.
% \newblock Degeneracy of the {$b$}-boundary in general relativity.
% \newblock {\em Comm. Math. Phys.}, 208(2):331--353, 1999.

\bibitem[Shi93]{Shi:93}
Katsuhiro Shiohama.
\newblock {\em An introduction to the geometry of {A}lexandrov spaces},
  volume~8 of {\em Lecture Notes Series}.
\newblock Seoul National University, Research Institute of Mathematics, Global
  Analysis Research Center, Seoul, 1993.

\bibitem[SKM{\etalchar{+}}03]{SK:03}
Hans Stephani, Dietrich Kramer, Malcolm MacCallum, Cornelius Hoenselaers and
  Eduard Herlt.
\newblock {\em Exact solutions of {E}instein's field equations}.
\newblock Cambridge Monographs on Mathematical Physics. Cambridge University
  Press, Cambridge, second edition, 2003.

\bibitem[SS18]{SS:18}
Clemens S\"{a}mann and Roland Steinbauer.
\newblock On geodesics in low regularity.
\newblock {\em Journal of Physics: Conference Series}, 968(1):012010, 2018.

\bibitem[SS21]{SS:21}
Benedict Schinnerl and Roland Steinbauer.
\newblock A note on the {G}annon-{L}ee theorem.
\newblock {\em Lett. Math. Phys.}, 111(6):Paper No. 142, 17, 2021.

\bibitem[Ste08]{Ste:08}
Roland Steinbauer.
\newblock A note on distributional semi-{R}iemannian geometry.
\newblock {\em Novi Sad J. Math.}, 38(3):189--199, 2008.

\bibitem[Ste14]{Ste:14}
Roland Steinbauer.
\newblock Every {L}ipschitz metric has {$C^1$}-geodesics.
\newblock {\em Classical Quantum Gravity}, 31(5):057001, 3, 2014.

\bibitem[SV09]{SV:09}
Roland Steinbauer and James~A. Vickers.
\newblock On the {G}eroch-{T}raschen class of metrics.
\newblock {\em Classical Quantum Gravity}, 26(6):065001, 19, 2009.

\bibitem[Vil09]{Vil:09}
C\'{e}dric Villani.
\newblock {\em Optimal transport}, volume 338 of {\em Grundlehren der
  mathematischen Wissenschaften [Fundamental Principles of Mathematical
  Sciences]}.
\newblock Springer, Berlin, 2009.
\newblock Old and new.

\bibitem[Vis95]{Vis:95}
Matt Visser.
\newblock {\em Lorentzian wormholes}.
\newblock AIP Series in Computational and Applied Mathematical Physics.
  American Institute of Physics, Woodbury, NY, 1995.
\newblock From Einstein to Hawking.

\end{thebibliography}

\newcommand{\etalchar}[1]{$^{#1}$}

\end{document}